\documentclass[aps,superscriptaddress,nofootinbib,floatfix]{revtex4-2}

\usepackage{epsfig}
\usepackage{bm}
\usepackage{amssymb}
\usepackage{amsmath}
\usepackage{color}
\usepackage{subfigure}
\usepackage{float}

\pdfoutput=1
\newcommand{\be}{\begin{equation}}
\newcommand{\ee}{\end{equation}}
\newcommand{\bea}{\begin{eqnarray}}
\newcommand{\eea}{\end{eqnarray}}
\newcommand{\bml}{\begin{subequations}}
\newcommand{\eml}{\end{subequations}}

\newcommand{\indep}{\perp \!\!\! \perp}

\begin{document}

\title{Relativistic Dissipative Magnetohydrodynamics from the Boltzmann equation for a two-component gas}
\date{\today}
\author{Khwahish Kushwah}
\affiliation{Instituto de F\'isica, Universidade Federal Fluminense (UFF), Niter\'oi,
24210-346, RJ, Brazil}
\email{khwahish\_kushwah@id.uff.br}

\author{Gabriel S.~Denicol}
\affiliation{Instituto de F\'isica, Universidade Federal Fluminense, UFF, Niter\'oi,
24210-346, RJ, Brazil}
\email{gsdenicol@id.uff.br}

\begin{abstract}
We derive the equations of motion of relativistic magnetohydrodynamics, as well as microscopic expressions for all of its transport coefficients, from the Boltzmann equation using the method of moments. In contrast to reference Phys.~Rev.~D 98(7) 2018, where a single component gas was considered, we perform our derivation for a locally neutral fluid composed of two massless particle species with opposite charges. We demonstrate that the magnetohydrodynamical equations of motion become dramatically different for this more realistic system. The shear-stress tensor no longer obeys a single differential equation; it breaks into three non-degenerate components with respect to the magnetic field, each evolving according to different dynamical equations. For large magnetic fields, we further show that the solution of this theory displays oscillatory behavior that can no longer be described by an Israel-Stewart-like theory. Finally, we investigate the derived equations in a Bjorken flow scenario.

\end{abstract}

\maketitle

\section{Introduction}

Relativistic magnetohydrodynamics (RMHD) is a theoretical framework that describes the dynamics of relativistic fluids in the presence of magnetic fields. Powerful magnetic fields are produced in nature and in experiment, playing crucial roles in high-energy heavy-ion collisions \cite{li2016electromagnetic, Hattori_2022, tuchin2013particle}, astrophysics \cite{rezzolla_book}, and the early universe \cite{ Brandenburg_1996}. For instance, in the early stages of heavy-ion collisions, nuclei beams generate intense magnetic fields, reaching peaks of $\sim 10^{19}$ gauss (RHIC) and $\sim 10^{20}$ gauss (LHC) \cite{Hattori_2017,Hattori_2022, Kharzeev_2008, SKOKOV_2009, Voronyuk_2011, Bzdak_2012, Zhong_2014, Holliday_2017,Dash:2023kvr}. These extreme electromagnetic fields that are created during the initial stages of heavy-ion collisions, may significantly impact the dynamics of the quark-gluon plasma (QGP), also formed at the early stages of the reaction \cite{wang2022incomplete,bloczynski2013azimuthally,tuchin2013time, deng2012event,yan2023dynamical, skokov2009estimate, voronyuk2011electromagnetic, bzdak2012event,Dash:2022xkz}. The framework of relativistic hydrodynamics \cite{rocha2023theories} is quite successful in explaining the dynamical evolution of heavy-ion collisions \cite{shen2020recent, Heinz_2013, GALE_2013} but, so far, these state of the art models do not include the effects of magnetic fields. Including these effects requires the derivation of causal relativistic magnetohydrodynamic equations, which do not display the unphysical features of relativistic Navier-Stokes theory \cite{pichon:65etude, hw1985}.

The derivation of causal formulations of relativistic magnetohydrodynamics  has been addressed by several authors \cite{Armas_2022,denicol2018nonresistive,denicol2019resistive,nonResistive_chapman, Resistive, hattori2016electrical, hattori2017bulk, hattori2017heavy, hattori2017longitudinal,li2018shear}. In principle, most candidates for relativistic magnetohydrodynamics correspond to extensions of the traditional Israel-Stewart theory \cite{Israel-stewart,ISRAEL,israel1979annals, israel1979jm} for plasmas, by coupling it to the Maxwell equations. In general, the presence of a magnetic field can considerably modify the structure of the equations of motion due to the spatial anisotropy introduced by the magnetic field \cite{Huang_2011, Hernandez_2017, gedalin1991relativistic, gedalin1995generally}: this anisotropy breaks down the degeneracy of several dissipative quantities, leading to the emergence of additional transport coefficients that display distinct values with respect to the direction of the magnetic field \cite{ghosh2022shear, nonResistive_chapman, Resistive, denicol2018nonresistive, denicol2019resistive, singh2020momentum, dash2020anisotropic, chen2020calculation, dey2021noninteracting}. In particular, relativistic magnetohydrodynamics has been derived in a kinetic theory framework \cite{denicol2018nonresistive, denicol2019resistive} using the traditional method of moments \cite{Denicol_Rischke}.

In the context of kinetic theory, magnetohydrodynamics has only been derived considering single-component gases \cite{denicol2018nonresistive, denicol2019resistive,Tinti:2018qfb,Resistive, nonResistive_chapman}. However, a fluid consisting of a single type of point-like charged particles is inherently unstable and incapable of reaching equilibrium. Consequently, such systems may lack the capacity to offer even a qualitative understanding of the problem. The goal of this paper is to address this severe limitation and derive magnetohydrodynamic from kinetic theory considering the simple, yet more realistic system, of a gas made of two massless particles species with opposite charges, with no dipole moment or spin. We find that the theory derived for this system is qualitatively different from those obtained for a single-component gas. In our case, different components of the shear-stress tensor with respect to the magnetic field obey distinct equations of motion -- a departure from the single-particle fluid scenario, where the presence of a magnetic field simply leads to the inclusion of additional transport coefficients \cite{denicol2018nonresistive, denicol2019resistive, Resistive, nonResistive_chapman}. Another distinctive aspect of our calculations is that, when the system is subjected to a relatively strong magnetic field, the shear-stress tensor displays oscillatory behavior as it approaches its  asymptotic equilibrium state. This is a drastic deviation from the typical exponential decay to equilibrium observed for the shear stress tensor in conventional hydrodynamic and magnetohydrodynamic approaches. Due to these oscillations, we find that the system can no longer be accurately described by a standard Israel-Stewart-like theory. 

This paper is organized as follows: in Sections \ref{sec:Boltzmann_equation} and \ref{sec:maxwells equation}, we discuss the Boltzmann equation in the presence of eletromagnetic fields. 
In \ref{sec: eom} we derive the fluid-dynamical equations in the presence of magnetic fields using the 14-moment approximation \cite{israel1979annals,Denicol_14_moment} and a power-counting scheme. In particular, in section \ref{sub:newprojections} we discuss the tensor decomposition of the shear-stress tensor with respect to the direction of the magnetic field. In section \ref{sec:Bjorken flow}, we solve the derived equations assuming the highly symmetric flow configuration given by the Bjorken flow \cite{Bjorken} and analyze the effects that the magnetic field can have on the dynamics of the shear-stress tensor. Lastly, in section \ref{sec:conclusion}, we summarize our results and make our concluding remarks.

\section{\label{sec:Boltzmann_equation}Boltzmann equation}
We consider a relativistic dilute gas of charged particles. The state of this system is described by the single particle momentum distribution function of each particle species, $f_\textbf{k}^i$, whose time evolution is described by the Boltzmann equation. The Boltzmann equation is an integro-differential equation of the following form \cite{DeGroot},
\begin{equation}
    k^{\mu}\partial_{\mu}f_\textbf{k}^i + qF^{\mu\nu}k_{\nu}\frac{\partial}{\partial k^{\mu}}f_\textbf{k}^i = \sum_j C[f_{\textbf{k}}^i, f_{\textbf{k}}^j],
\end{equation}
where $C[f_{\textbf{k}}^i, f_{\textbf{k}}^j]$ is the collision term, $F^{\mu\nu}$ is the electromagnetic field tensor and $k^\mu$ is the particle 4-momentum. The collision term is non-linear and contains integrals over the momentum of the distribution function of all particle species, rendering the equation challenging to solve.

The energy-momentum tensor and net-charge (electric) four-current are expressed as the following momentum integrals of the single-particle distribution function 
\begin{equation*}
T^{\mu \nu } = \sum_{i}\left\langle k^{\mu}_ik^{\nu }_i\right\rangle_i \equiv \sum_i T^{\mu\nu}_i ,\quad N^{\mu
}=\sum_{i}q_i\left\langle k^{\mu}_i\right\rangle_i \equiv \sum_i N^{\mu}_i
\end{equation*}%
where the summations above are over all particle species and we have used the following notation, 
\begin{equation}
\left\langle \ldots \right\rangle_i \equiv g\int \frac{d^{3}\mathbf{k}}{(2\pi )^{3}k^{0}%
}(\ldots )f_{\mathbf{k}}^i.
\end{equation}%
Here, $g$ is the degeneracy factor and $\displaystyle{ k^0 = \sqrt{\mathbf{k}^2 + m_0^2}}$ is the on-shell energy. These current are associated to conserved quantities and satisfy the
continuity equations (in the absence of electromagnetic fields), 
\begin{equation}
\label{bla}
    \partial _{\mu }T^{\mu \nu }=0; \qquad \partial _{\mu
}N^{\mu }=0.
\end{equation}

It is convenient to decompose $T^{\mu \nu }$ and $N^{\mu }$ in terms of the fluid's collective 4-velocity field, $u^{\mu }$. These currents are then re-expressed as \cite{DeGroot, barrow2007cosmology, bekenstein1978new}
\begin{equation}
\begin{split}
T^{\mu \nu } &= \epsilon u^{\mu }u^{\nu }-\Delta ^{\mu \nu }\left( P_{0}+\Pi
\right)+\pi ^{\mu \nu }, \\
N^{\mu } &= nu^{\mu }+V^{\mu },  \label{Nmuformel}
\end{split}
\end{equation}
where we introduced the energy density $\epsilon $, the thermodynamic pressure $P_{0}$, the bulk viscous pressure $\Pi $, the shear-stress tensor $\pi ^{\mu \nu }$, the net-charge density $n$, and the net-charge diffusion current $V^{\mu }$. We also defined the spatial projector $\Delta ^{\mu \nu }=g^{\mu \nu }-u^{\mu }u^{\nu }$ and employed Landau's
definition \cite{landau1987fluid} of the fluid velocity as an eigenvector of $T^{\mu \nu }$ with eigenvalue $\epsilon $, that is, $T^{\mu \nu }u_{\nu }=\epsilon u^{\mu }$.
In this scheme, each new variable introduced is expressed by a given contraction/projection of the currents with $u^{\mu }$ and $\Delta ^{\mu \nu}$ \cite{Israel-stewart}, 
\begin{eqnarray*}
\epsilon &\equiv&u_{\mu }u_{\nu }T^{\mu \nu }\text{,}\qquad P_{0}+\Pi \equiv-\frac{1}{3}
\Delta _{\mu \nu }T^{\mu \nu }\text{,} \qquad \pi ^{\mu \nu } \equiv  %\Delta _{\alpha \beta }^{\mu \nu}
T^{\langle\mu\nu\rangle }\text{,}\\  n &\equiv& u_{\mu }N^{\mu }\text{, } \hspace{4.2 em} V^{\mu }\equiv N^{\left\langle \mu \right\rangle }.
\end{eqnarray*}
For convenience, we adopt the notation 
\begin{align*}
A^{\left\langle \mu \right\rangle
}& \equiv \Delta _{\nu }^{\mu }A^{\nu },\\
A^{\left\langle \mu \nu
\right\rangle }& \equiv \Delta _{\alpha \beta }^{\mu \nu }A^{\alpha \beta}.
\end{align*}
The latter definition used the double, traceless, symmetric projection operator,
\begin{equation}
    \Delta _{\alpha \beta }^{\mu \nu }=\frac{1}{2}\left(\Delta _{\alpha }^{\mu }\Delta
_{\beta }^{\nu }+\Delta _{\alpha }^{\nu }\Delta _{\beta }^{\mu }-\frac{2}{3}\Delta
^{\mu \nu }\Delta _{\alpha \beta }\right).
\end{equation}
Since our goal requires focusing on the equation of motion for the shear-stress tensor, most of the dissipative currents introduced above will play no role in our calculation. Nevertheless, we introduced them above for the sake of completeness.

\section{The Maxwell's equation and the Electromagnetic tensor}
\label{sec:maxwells equation}
The evolution of Electric and magnetic fields are given by Maxwell's equations,
\begin{equation}
    \begin{split}
        \partial_{\mu}F^{\mu\nu} & =N^{\nu}, \\
        \partial_{\mu}\Tilde{F}^{\mu\nu} & =0,
    \end{split}
\end{equation}
where the Faraday tensor, $F^{\mu\nu}$, is decomposed with respect to the fluid velocity in the following form \cite{cercignani2002relativistic, barrow2007cosmology}
\begin{equation}\label{faraday}
F^{\mu \nu }=E^{\mu }u^{\nu }-E^{\nu }u^{\mu }+\epsilon ^{\mu \nu \alpha \beta } u_{\alpha }B_{\beta},
\end{equation}
and its Hodge dual is
\begin{equation}\label{hodgedual}
\Tilde{F}^{\mu \nu }=\frac{1}{2}\epsilon^{\mu\nu\alpha\beta}F_{\alpha\beta} = B^{\mu }u^{\nu}-B^{\nu}u^{\mu}-\epsilon^{\mu \nu
\alpha \beta } u_{\alpha }E_{\beta}.
\end{equation}
This rank-two, antisymmetric tensor is composed of the electric field 4-vector, $E^{\mu }$, and the magnetic field 4-vector, $B^{\mu }$, which are orthogonal to the fluid 4-velocity, $ E^{\mu }u_{\mu }=0$ and $B^{\mu }u_{\mu }= 0$. Moreover, in the local rest frame
of the fluid, these 4-vectors coincide with usual electric and magnetic fields, i.e., $E^{\mu}_{LR} = ( 0, \Vec{E})^T$ and $B^{\mu}_{LR} = ( 0, \Vec{B})^T$ with the following definitions,
\begin{equation*}
E^{i}  = F^{i0},  \qquad
B^{i} =-\frac{1}{2}\epsilon^{ijk}F_{jk}.
\end{equation*}%
In a covariant form, we also have
\begin{equation*}
E^{\mu } =u_{\nu }F^{\mu \nu }, \qquad
B^{\mu } =\frac{1}{2}\epsilon ^{\mu \nu \alpha \beta }F_{\nu \alpha}u_{\beta}.
\end{equation*}
Further, $N^{\mu}$ is accurately characterized by Eq.\eqref{Nmuformel} as the electric charge four-current. It also serves as a source for the electromagnetic field. The conservation of the total energy and momentum of the gas is disrupted by the presence of the fields, which can exchange energy and momentum with the system. The conservation laws \eqref{bla} are now re-expressed as,
\begin{equation}
    \partial_{\mu}T^{\mu\nu} = -F^{\nu\lambda}N_{\lambda}. 
\end{equation}
Naturally, the electric charge four-current of the fluid remains conserved, i.e., $\displaystyle{\partial_\mu N^{\mu} = 0}$. For simplicity, we shall consider a locally neutral fluid and will disregard any contribution from the electric field 4-vector, $E^\mu$, throughout this work. In absence of $E^\mu$, Eqs.\eqref{faraday} and \eqref{hodgedual} simplify as follows 
\begin{align}
  F^{\mu\nu} & \rightarrow B^{\mu\nu} = \epsilon^{\mu\nu\alpha\beta}u_{\alpha}B_\beta, \\
  \Tilde{F}^{\mu\nu}& \rightarrow \Tilde{B}^{\mu\nu} = B^{\mu}u^\nu-B^{\nu}u^\mu.
\end{align}

\section{Equations of motion}
\label{sec: eom}

 Our goal is to find the equations of motion for the shear-stress tensor, $\pi^{\mu\nu}$, in the presence of a moderately large magnetic field. We consider a locally neutral fluid composed of two types of massless classical particles with opposite electric charges and vanishing dipole moment or spin, so that the fluid has vanishing magnetization and polarization. For this system the Boltzmann equation reads,
\begin{subequations}
    \begin{align}
        k^{\mu} \partial_\mu f^{-}_{k} + q^{-} k_{\nu} F^{\mu \nu} \frac{\partial}{\partial k^{\mu}}f_k^{-} & = C[f^{-},f^+],
        \label{minus}\\
        k^{\mu} \partial_\mu f^{+}_{k} + q^{+} k_{\nu} F^{\mu \nu} \frac{\partial}{\partial k^{\mu}}f_k^+ & =C[f^+,f^{-}],
        \label{plus}
    \end{align}
\end{subequations}\normalsize
where we consider only elastic collisions (for supporting derivation see \cite{DNMR}), 
\small\begin{align}
    C[f^-,f^+]&\equiv \frac{1}{2} \int dK^{'} dP dP^{'} W^{--}_{KK^{'} \leftrightarrow PP^{'}} \left( f_p^- f_{p^{'}}^- - f_k^- f_{k^{'}}^- \right) + \int dK^{'} dP dP^{'} W^{-+}_{KK^{'} \leftrightarrow PP^{'}} \left( f_p^- f_{p^{'}}^+ - f_k^- f_{k^{'}}^+ \right),\\
   C[f^+,f^-] & \equiv \frac{1}{2} \int  dK^{'} dP dP^{'} W^{++}_{KK^{'} \leftrightarrow PP^{'}} \left( f_p^+ f_{p^{'}}^+ - f_k^+ f_{k^{'}}^+ \right) + \int  dK^{'} dP dP^{'} W^{-+}_{KK^{'} \leftrightarrow PP^{'}} \left( f_p^+ f_{p^{'}}^- - f_k^+ f_{k^{'}}^- \right). 
\end{align}
In this section, the index $\pm$ refers to the respective particle species. The transition rate can be defined in terms of the total cross section, $\sigma_T$, as \cite{DeGroot}
\begin{equation}
   W_{kk^{'}\rightarrow pp^{'}}^{+-} = s\sigma_T^{+-} (2\pi)^5 \delta^{(4)}( k^{\mu} + k^{'\mu} - p^{\mu} - p^{'\mu}). 
\end{equation}
Above, we assumed that the cross-sections are constant. In the following, we shall further assume that, $\sigma^{++}_T = \sigma^{--}_T \equiv \sigma_T$, which will simplify the derivation of the fluid dynamical equations. 

For the sake of convenience, we introduce the energy momentum tensor, $T^{\mu\nu}_{\pm}$, and the net-charge current, $N^{\mu}_{\pm}$,  of each particle species. As done in section \ref{sec:Boltzmann_equation}, we decompose these tensors in terms of the fluid 4-velocity, 
\begin{equation}
\begin{split}
T^{\mu \nu }_{\pm} &= \epsilon^{\pm} u^{\mu }u^{\nu }-\Delta ^{\mu \nu } P^{\pm} +h^{\mu}_{\pm}u^{\nu}+h^{\nu}_{\pm}u^{\mu}
+\pi ^{\mu \nu }_{\pm}, \\
N^{\mu}_{\pm} &= n^{\pm}u^{\mu }+V^{\mu }_{\pm}. 
\end{split}
\end{equation}
Here we have used the same notation as above, with the indices '$\pm$' indicating the respective particle species. Therefore, $\epsilon^{\pm} $, $P^{\pm}$, 
$h^{\mu}_{\pm}$, $\pi^{\mu\nu}_{\pm}$, $n^{\pm}$ and $V^{\mu}_{\pm}$ are the energy density, the isotropic
pressure, the energy diffusion 4-current, the shear-stress tensor, the net-charge density , and the net-charge diffusion 4-current for the corresponding $\pm$ particle species, respectively.
In this scheme, each new variable introduced is expressed by a given contraction/projection of the currents with $u^{\mu}$ and $\Delta ^{\mu \nu}$, 
\begin{eqnarray*}
\epsilon^{\pm} &\equiv&u_{\mu }u_{\nu }T^{\mu \nu }_{\pm}\text{,}\qquad P^{\pm}+\Pi^{\pm} \equiv-\frac{1}{3}%
\Delta _{\mu \nu }T^{\mu \nu}_{\pm}\text{,} \qquad \pi ^{\mu \nu }_{\pm} \equiv  
T^{\langle\mu\nu\rangle }_{\pm}\text{,}\\  h^{\mu}_{\pm} &\equiv& u_{\alpha}T^{\langle\mu\rangle\alpha}_{\pm}  \text{, } \hspace{4.2 em} n^{\pm} \equiv u_{\mu }N^{\mu }_{\pm}\text{, } \hspace{4.2 em} V^{\mu }_{\pm}\equiv N^{\left\langle \mu
\right\rangle }_{\pm}.
\end{eqnarray*}

% In eq. \eqref{landau}, we used the traditional definition of Landau frame considering total $T^{\mu\nu}$ of the system since we simply can not write the same equation for individual particle species.

\subsection{Matching conditions}
\label{sub:matching conditions}
We now introduce a reference local equilibrium state and decompose the energy density, isotropic pressure and the net-charge density (for 'plus' and 'minus' particles) as
\begin{equation}
    \begin{split}
        \epsilon^{\pm} & \equiv \epsilon_0^{\pm}+\delta\epsilon^{\pm},\\
        P^{\pm} & \equiv P_0^{\pm} + \Pi^{\pm}, \\
        n^{\pm} & \equiv n_0^{\pm} +\delta n^{\pm},
    \end{split}
\end{equation}
where $\epsilon_0^{\pm}$, $P_0^{\pm}$, and $n_0^{\pm}$ are energy density, pressure, and net-charge density of the '$\pm$' particles of the system at equilibrium, respectively. The corresponding $\delta\epsilon^{\pm}$ and $\delta n^{\pm}$ are non-equilibrium corrections to the energy density, and net-charge density, respectively, for '$\pm$' particles and $\Pi^{\pm}$ is the bulk viscous pressure. 
We note that since we are considering a system composed of massless particles, $\delta\epsilon^{\pm} = 3\Pi^{\pm}$.

In this work, we impose Landau matching conditions which fixes the 4-velocity as an eigenvector of $T^{\mu\nu}$ in such a way that
\begin{align}
 \label{landau}
    u_\mu \left(T^{\mu\nu}_+ + T^{\mu\nu}_-\right) & = \epsilon_0 (\mu,T) u^\nu,\\
    u_\mu \left(N^{\mu}_+ - N^{\mu}_-\right) & = n_0(\mu,T).
\end{align}
This implies that the total energy density and, in our case, the electric net-charge density are given by their respective equilibrium value,

\begin{align}
\epsilon^+ + \epsilon^- & \equiv \epsilon_0(\mu,T) \Longrightarrow \delta\epsilon^+ + \delta\epsilon^- = 0, \label{mtchcondition1}\\
 n^{+}- n^{-} & \equiv n_0(\mu,T) \Longrightarrow \delta n^+ - \delta n^- = 0.\label{mtchcondition2}
\end{align}
Finally, Eq.~\eqref{landau} imposes that the total energy diffusion 4-current vanishes, $$h^{\mu}_{+} + h^{\mu}_{-} = 0,$$
nevertheless the energy diffusion 4-current of individual particle species does not necessarily vanish.

Equations \eqref{mtchcondition1}  and \eqref{mtchcondition2} define the electric charge chemical potential, $\mu$, and a local temperature, $T$. This matching condition will guarantee that the bulk viscous pressure completely vanishes. We note that the dissipative currents of each species $\delta\epsilon^{\pm}$, $\Pi^{\pm}$ and $\delta n^{\pm}$ are not necessarily zero. Nevertheless since we are considering massless particles, these fields will be at least of second order in gradients and thus, will be neglected in our derivation \cite{deBrito:2023tgb}. 

Finally, we further assume that the chemical potential is zero, 
\begin{equation}
    \mu = 0 \Longrightarrow n_0(T) = 0.
\end{equation}
This assumption also implies that the net-charge diffusion 4-current can be neglected. The energy diffusion 4-current of each particle species will not disappear, but will become at least second order in gradients and, thus, will also be neglected in our calculations. Consequently, we are only required to derive the equations of motion for the shear-stress tensor.

\subsection{Exact equations of motion}
\label{sub:exact eom}
We directly calculate the time derivative of the shear stress tensor of each particle species following the procedure outlined in Ref.~\cite{DNMR},
\begin{equation}\label{pi1}
    \Dot{\pi}^{\mu\nu}_{\pm} \equiv \frac{d}{d \tau}\pi^{\mu\nu}_{\pm} = \frac{d}{d \tau}\int dk \ k^{\langle\mu} k^{\nu\rangle} f^{\pm}_k .
\end{equation}
%  Non-equilibrium corrections to the hydrodynamical quantities can be expressed in terms of these irreducible moments 
% \begin{equation}
%     \begin{split}
%         \delta n_{\pm} & \equiv \delta N^{\mu}_{\pm} u_{\mu} = \rho_1^{\pm}\\
%         \delta\epsilon_{\pm} & \equiv \delta T^{\mu\nu}_{\pm} u_{\mu}u_{\nu} = \rho_2^{\pm}\\
%         \pi^{\mu\nu}_{\pm} & \equiv \Delta^{\mu\nu}_{\alpha\beta}\delta T^{\alpha\beta}_{\pm} = \rho^{\mu\nu}_{0\pm}
%     \end{split}
% \end{equation}
For a fluid consisting of two particle species, equation (\ref{pi1}) can be re-expressed as 
\begin{equation}\label{pi}
\begin{split}
\Delta^{\mu\nu}_{\alpha\beta}\Dot{\pi}^{\alpha\beta}_+= \Dot{\pi}^{\langle\mu\nu\rangle}_+ & = \int dk \ k^{\langle\mu}k^{\nu\rangle}
    \frac{d}{d \tau} f^+ - 2 \Dot{u}^{\langle\mu}h^{\nu\rangle}_+,\\
\Delta^{\mu\nu}_{\alpha\beta}\Dot{\pi}^{\alpha\beta}_-= \Dot{\pi}^{\langle\mu\nu\rangle}_- &=  \int dk \ k^{\langle\mu}k^{\nu\rangle}
    \frac{d}{d \tau}f^- - 2 \Dot{u}^{\langle\mu}h^{\nu\rangle}_{-},
    \end{split}
    \end{equation}
where the comoving derivative of the single particle distribution function is calculated from the Boltzmann equation in the following form
\begin{equation}
    \label{boltzeq2} 
   E_k\frac{d}{d\tau}f^{\pm}  = - k^{\mu} \nabla_{\mu} f^{\pm} \mp |q| k_{\nu} B^{\mu \nu} \frac{\partial}{\partial k^{\mu}}f^{\pm}  + C[f^+, f^-].
\end{equation}
Here we defined $\displaystyle{E_k = k^\mu u_\mu}$, the energy of a particle in the local rest frame of the fluid. By replacing Eq.~(\ref{boltzeq2}) into Eq.~(\ref{pi}), we obtain an equation for the shear-stress tensor of each particle species \footnote{Analogous to the equations derived in \cite{denicol2018nonresistive} for a single-component gas.},
\begin{equation}
    \begin{split}\label{piequation}
    \Dot{\pi}_+^{\langle\mu\nu\rangle}=& -\Delta^{\mu\nu}_{\alpha\beta}\nabla_\lambda \langle E_k^{-1} k^{\langle \alpha} k^\beta k^{\lambda\rangle}\rangle_+ + \frac{2}{5}\Delta^{\mu\nu}_{\alpha\beta}\nabla^\alpha \langle E_k^{-1} k^{\langle\beta\rangle} \rangle_+ -\Delta^{\mu\nu}_{\alpha\beta} \nabla_\lambda u_\kappa \langle E_k^{-2} k^{\langle\alpha} k^{\beta} k^{\lambda}k^{\kappa\rangle} \rangle_+ - \frac{4}{3}\pi^{\langle\mu\nu\rangle}_{+}\theta
        +\frac{8}{15} \sigma^{\mu\nu}  \epsilon^+ \\ & - \frac{10}{7}\Delta^{\mu\nu}_{\alpha\beta}\sigma^{\alpha}_{\lambda}\pi^{\beta\lambda}_{+}-  2\Delta^{\mu\nu}_{\alpha\beta}\omega^{\ \ \beta}_{\lambda}\pi^{\alpha\lambda}_+  -2|q|\Delta^{\mu\nu}_{\alpha\beta}Bb_{\lambda}\ ^{\alpha} \langle E_k^{-1} k^{\langle\beta}k^{\lambda\rangle} \rangle_+ +2|q| E^{\langle \mu}n^{\nu\rangle}_+ - 2 \Dot{u}^{\langle\mu}h^{\nu\rangle}_{+} + C^{\langle\mu\nu\rangle}_+ ,\\
        \Dot{\pi}_-^{\langle\mu\nu\rangle}  = & -\Delta^{\mu\nu}_{\alpha\beta}\nabla_\lambda \langle E_k^{-1} k^{\langle \alpha} k^\beta k^{\lambda\rangle}\rangle_- + \frac{2}{5}\Delta^{\mu\nu}_{\alpha\beta}\nabla^\alpha \langle E_k^{-1} k^{\langle\beta\rangle} \rangle_- -\Delta^{\mu\nu}_{\alpha\beta} \nabla_\lambda u_\kappa \langle E_k^{-2} k^{\langle \alpha} k^{\beta} k^{\lambda}k^{\kappa\rangle} \rangle_- - \frac{4}{3}\pi_-^{\langle\mu\nu\rangle}\theta +\frac{8}{15} \sigma^{\mu\nu} \epsilon^- \\ &- \frac{10}{7}\Delta^{\mu\nu}_{\alpha\beta}\sigma^{\alpha}_{\lambda}\pi_-^{\beta\lambda}-  2\Delta^{\mu\nu}_{\alpha\beta}\omega^{\ \ \beta}_{\lambda}\pi^{\alpha\lambda}_- +2|q|\Delta^{\mu\nu}_{\alpha\beta}Bb_{\lambda}\ ^{\alpha}\langle E_k^{-1} k^{\langle\beta}k^{\lambda\rangle} \rangle_- - 2|q| E^{\langle \mu}n^{\nu\rangle}_- - 2 \Dot{u}^{\langle\mu}h^{\nu\rangle}_{-} + C^{\langle\mu\nu\rangle}_-,
        %- \frac{\pi^{\langle\mu\nu\rangle}_- }{5 } \left(  3n^-\sigma_T^{--} + 4\ n^+ \sigma_T^{+-} \right) + \frac{\pi^{\langle\mu\nu\rangle}_+}{5}  n^-\sigma_T^{+-} 
    \end{split}
\end{equation}
where we have introduced a dimensionless antisymmetric tensor $b^{\mu\nu}$ $\equiv$ $- B^{\mu\nu}/B $, with $B^{\mu\nu}B_{\mu\nu}$ = 2$B^2$, and the irreducible tensors $k^{\langle\mu_1} \cdots k^{\mu_\ell \rangle} \equiv \Delta^{\mu_1 \ldots \mu_\ell}_{\nu_1 \ldots \nu_\ell}k^{\nu_1} \cdots k^{\nu_\ell}$, with $\Delta^{\mu_1 \ldots \mu_\ell}_{\nu_1 \ldots \nu_\ell}$ being a $2\ell$-rank symmetric and traceless projection operator orthogonal to $u^\mu$ \cite{DeGroot,Denicol_Rischke}. Additionally, we have also defined the shear tensor, $\displaystyle{\sigma^{\mu\nu} \equiv \nabla^{\langle\mu} \ u^{\nu\rangle}}$, the expansion scalar, $\displaystyle{\theta \equiv \nabla_\mu u^\mu}$ and the vorticity tensor, $\displaystyle{\omega = (\nabla^{\mu}u^\nu - \nabla^\nu u^\mu)/2}$. Finally, we defined the following moments of the collision term, 
\begin{align}
    C^{\langle\mu\nu \rangle}_+ & = \int dK E_k^{-1} k^{\langle\mu}k^{\nu\rangle} C[f^{+}, f^-], \\
    C^{\langle\mu\nu\rangle}_- & = \int dK E_k^{-1} k^{\langle\mu}k^{\nu\rangle} C[f^{-}, f^{+}].
\end{align}
The equations derived above are exact but are not closed in terms of the fluid-dynamical fields. In order to obtain a closed set of equations we need to impose some approximations that will simplify the moments of the collision term and the term that couples these moment equations with the magnetic field.

\subsection{14 moment approximation}
\label{sub:14moment}
In this section, we discuss the 14-moment approximation for the single-particle distribution function of each species, which will be used to close the moments equations \eqref{minus} and \eqref{plus}. We follow the original procedure constructed by Israel and Stewart \cite{israel1979annals, israel1979jm} and express the single-particle distribution function of each particle species as,
\begin{equation}
    f_k^{\pm} = \exp({y_{\mathbf{k}}^{\pm}}).
\end{equation}
Next, the field $y_{\mathbf{k}}$ is expanded in momentum space around its local-equilibrium value, $y_{0\mathbf{k}} = \alpha-\beta u_{\mu}k^{\mu}$, with $\alpha=\mu/T$ being the thermal potential and $\beta=1/T$ the inverse temperature. This is a series in terms of Lorentz-tensors formed from 4-momentum $k^{\mu}$, 
\begin{equation}
    \delta y_{\mathbf{k}}^{\pm} \equiv y_{\mathbf{k}} - y_{0\mathbf{k}} = \varepsilon^{\pm} + k^{\mu}\varepsilon_{\mu}^{\pm}+k^{\mu}k^{\nu}\varepsilon_{\mu\nu}^{\pm}+k^{\mu}k^{\nu}k^{\lambda}\epsilon_{\mu\nu\lambda}^{\pm}+ \cdots.
\end{equation}
To first order in $\delta y_{\mathbf{k}}$, we obtain
\begin{equation}
     f_k^{\pm} = f_{0\mathbf{k}}^{\pm} + f_{0\mathbf{k}}^{\pm}%\Tilde{f}_{0\mathbf{k}}
     \delta y_{\mathbf{k}}^{\pm} + \mathcal{O}\left(\delta y_{\mathbf{k}}^{2}\right).
\end{equation}

In the 14-moment approximation, the expansion of the nonequilibrium correction $\delta y_{\mathbf{k}}^{\pm}$ in powers of 4-momentum is truncated at second-order \cite{israel1979annals,Denicol_14_moment}.
That is, we only include the tensors 1, $k^{\mu}$, and $k^{\mu}k^{\nu}$ in the expansion, 
\begin{equation}
    \delta y_{\mathbf{k}}^{\pm} \approx \varepsilon^{\pm} + k^{\mu}\varepsilon_{\mu}^{\pm} + k^{\mu}k^{\nu}\varepsilon_{\mu\nu}^{\pm}.
\end{equation}
Without loss of generality, we can assume $\varepsilon_{\mu\nu}^{\pm}$ to be symmetric and traceless \footnote{The trace of $\varepsilon_{\mu\nu}^{\pm}$ can always be incorporated into the scalar expansion coefficient $\varepsilon^{\pm}$.}, thus leaving only 14 independent degrees of freedom in the expansion coefficients $\varepsilon^{\pm}$, $\varepsilon_{\mu}^{\pm}$, and $\varepsilon_{\mu\nu}^{\pm}$. These 14 degrees of freedom are usually matched to the degrees of freedom of $N^{\mu}_{\pm}$ and $T^{\mu\nu}_{\pm}$. Here, since we consider a system of massless particles and vanishing chemical potential, all scalar and 4-vector dissipative currents either vanish or are of a higher order, and thus are not considered in our analyses. This implies that only $\varepsilon_{\mu\nu}^{\pm}$ is not zero in the truncated expansion above and this coefficient can be directly matched to $\pi^{\mu\nu}_{\pm}$. This leads to the well-known expression \cite{israel1979annals,Denicol_14_moment},
\begin{equation}
    \delta f^{\pm} = \frac{f_0 \left(\pi^{\pm}_{\mu\nu}\ k^{\mu}k^{\nu} \right)}{2\left(\epsilon^{\pm}_0+P_0^{\pm} \right)T^2}.
\end{equation}
When expanding the distribution function in powers of \(\delta y_{\textbf{k}}\), only the leading term of the expansion was retained. In order to be consistent, the same approximation must be applied to the collision term and all terms of order \(\mathcal{O}(\delta y^2_{\textbf{k}})\) and higher are omitted. Then,  using the 14-moment approximation, the collision term takes the following form 
\begin{equation}
\begin{split}
   C^{\mu\nu}_- =& -\frac{6}{5} \sigma_T^{--} n_0^-\pi^{\mu\nu}_- + \frac{2}{5} \sigma_T^{+-}\left( n_0^- \pi_+^{\mu\nu} - 4n_0^+\pi_-^{\mu\nu} \right) ,\\
   C^{\mu\nu}_+ =& -\frac{6}{5} \sigma_T^{++}n_0^+\pi^{\mu\nu}_+ + \frac{2}{5}\sigma_T^{-+} \left( n_0^+ \pi_-^{\mu\nu} - 4n_0^-\pi_+^{\mu\nu} \right) ,
    \end{split}
    \end{equation}
where $n_0^{+}$ and $n_0^{-}$ correspond to the particle densities of specie 'plus' and 'minus' in equilibrium, respectively. Since we assume that the electric-charge chemical potential vanishes, $n_0^{+}=n_0^{-}\equiv \hat{n}_0$. Using the 14-moment approximation, we can also demonstrate that,
\begin{equation}
\begin{split}
     \langle E_k^{r} k^{\langle\mu\rangle} \rangle_{\pm} &= 0 ,\\
     \langle E_k^{-1} k^{\langle\beta}k^{\lambda\rangle} \rangle_{\pm} &= \frac{\pi^{\beta\lambda}_{\pm}}{5T} , \\  
    \langle E_k^{-1} k^{\langle \alpha} k^\beta k^{\lambda\rangle}\rangle_\pm &= 0, \\
   \langle E_k^{-2} k^{\langle\alpha} k^{\beta} k^{\lambda}k^{\kappa\rangle} \rangle_\pm &= 0.
\end{split}   
\end{equation}
In the first equation, $r$ is an arbitrary constant that satisfies, $r>-2$. 

We then replace these results into Eqs.\ \eqref{piequation} and rewrite them in terms of the total shear stress tensor, $\pi^{\mu\nu}=\pi_+^{\mu\nu}+ \pi_-^{\mu\nu}$, and a relative shear-stress tensor, $\delta\pi^{\mu\nu}\equiv \pi_+^{\mu\nu}- \pi_-^{\mu\nu}$. The result is, \begin{align}
\label{totalshear}
\Delta _{\alpha \beta }^{\mu \nu }\dot{\pi}^{\alpha \beta
}+\Sigma\pi^{\mu \nu }+\frac{2|q|B}{5T}b^{\lambda
\langle \mu}\delta\pi_{\lambda }^{\nu \rangle
}&=\frac{8}{15}\epsilon\sigma ^{\mu \nu }-\frac{4}{3}\pi^{\mu
\nu }\theta -\frac{10}{7}\sigma ^{\lambda \langle \mu }\pi
_{\lambda }^{ \nu \rangle} -2\omega^{\lambda\langle\nu}\pi^{\mu \rangle}_{\lambda},\\
\label{deltashear}
\Delta _{\alpha \beta }^{\mu \nu }\delta\dot{\pi}^{\alpha \beta
}+\Sigma^{'}\delta\pi^{\mu \nu }+\frac{2|q|B}{5T}b^{\lambda
\langle \mu}\pi_{\lambda }^{\nu \rangle} &= -\frac{4}{3}\delta\pi^{\mu
\nu }\theta -\frac{10}{7}\sigma ^{\lambda \langle \mu}\delta\pi
_{\lambda }^{\nu\rangle} -2\omega^{\lambda\langle\nu}\delta\pi^{\mu \rangle}_{\lambda}.\hspace{1.2 cm}
\end{align}
Above, we defined two transport coefficients, 
\begin{equation}
    \Sigma =\frac{3\hat{n}_0}{5}\left(\sigma_T^{+-}+\sigma_T\right) \quad\text{and}\qquad \Sigma^{'}=\frac{\hat{n}_0}{5}\left(5\sigma_T^{+-}+3\sigma_T\right).
\end{equation} 
In the absence of a magnetic field, $\Sigma$ can be immediately identified as the inverse relaxation time of this fluid and can be related to the shear viscosity, $\eta$, as,
\begin{equation}
    \Sigma = \frac{1}{\tau_\pi} = \frac{\epsilon + P_0}{5\eta}.
    \label{shear}
\end{equation}
We observe that the equations of motion for $\pi^{\mu\nu}$ and $\delta\pi^{\mu\nu}$are coupled due to the presence of a magnetic field \footnote{This is due to our assumption $\sigma_T^{++} = \sigma_T^{--}$.}. A closed set of equations for $\pi^{\mu\nu}$ can be obtained in some limits, as will be discussed in the remainder of this paper. Nevertheless, before we implement this procedure, we shall discuss how the dissipative currents can be tensor-decomposed with respect to the magnetic field --  a procedure that will be required in order to simplify the coupling term that appeared in equations above by the existence of a magnetic field. 

\subsection{New Projections and definitions}
\label{sub:newprojections}

We introduce a normalized 4-vector, $b^{\mu}$, defined such that%
\begin{equation*}
b^{\mu }\equiv \frac{
B^{\mu }}{B}, \quad \text{where}\quad
-B_{\mu }B^{\mu }\equiv B^{2}\ \  \Longrightarrow \ \ b_{\mu }b^{\mu }=-1.
\end{equation*}%
Now we proceed to decompose a traceless second-rank tensor that is orthogonal to $u^\mu$, e.g. the shear-stress tensor $\pi^{\mu\nu}$, with respect to the direction of the magnetic field, $b^\mu$. That is,
\begin{equation}
\begin{split}
\label{projmag}
\pi ^{\mu \nu }=\pi _{\parallel }\left( b^{\mu }b^{\nu }+\frac{1}{2}\Xi
^{\mu \nu }\right) +2\pi _{\perp }^{\left( \mu \right. }b^{\left. \nu
\right) }+\pi _{\indep }^{\mu \nu },\hspace{5em}\\
\text{here}, \qquad \pi_{\parallel }\equiv b_{\alpha }b_{\beta }\pi^{\alpha\beta },\qquad \pi_{\perp
}^{\mu }\equiv -\Xi_{\alpha }^{\mu }b_{\beta }\pi^{\alpha\beta}, \qquad\pi
_{\indep}^{\mu \nu }\equiv \Xi_{\alpha \beta }^{\mu \nu }\pi ^{\alpha \beta
},
\end{split}
\end{equation}
where we defined the following projection operators onto the subspace
orthogonal to $u^{\mu }$ and $b^{\mu }$,
\begin{align}
\Xi ^{\mu \nu } &\equiv g^{\mu \nu }-u^{\mu }u^{\nu }+b^{\mu }b^{\nu } = \Delta^{\mu\nu} +b^{\mu}b^{\nu}, \\
\Xi _{\alpha \beta }^{\mu \nu } &\equiv \frac{1}{2}\left( \Xi _{\alpha
}^{\mu }\Xi _{\beta }^{\nu }+\Xi _{\beta }^{\mu }\Xi _{\alpha }^{\nu }-\Xi
^{\mu \nu }\Xi _{\alpha \beta }\right) .
\end{align}
For the sake of convenience, we further parameterize%
\begin{eqnarray}
b^{\mu \nu } &\equiv & -\epsilon ^{\mu \nu \alpha \beta }u_{\alpha}b_{\beta }.
\end{eqnarray}
Lets now consider the following complete, normalized, and orthogonal basis
$(u^{\mu },\ b^{\mu },\ x^{\mu },\ y^{\mu })$
in such a way that, in the local rest frame, if we define $b^{\mu}$ to be
in the longitudinal direction, we have that 
\begin{align}
u^{\mu } =\left( 1,0,0,0\right) ,  \qquad
x^{\mu } =\left( 0,1,0,0\right) , & \qquad
y^{\mu } =\left( 0,0,1,0\right) ,  \qquad
b^{\mu } =\left( 0,0,0,1\right) .
\end{align}
Or, in other words, $x^{\mu }$ and $y^{\mu }$ describe the plane orthogonal
to the magnetic field in the local rest frame of the fluid. For the sake of convenience, we further define a new basis 4-vector,%
\begin{equation}
\ell _{\pm }^{\mu }\equiv \frac{1}{\sqrt{2}}\left( x^{\mu }\pm iy^{\mu }\right) ,
\end{equation}%
which satisfy the conditions,
\begin{eqnarray}
\ell _{\mu }^{\pm }\ell _{\pm }^{\mu } &=&\frac{1}{2}\left( x_{\mu }\pm
iy_{\mu }\right) \left( x^{\mu }\pm iy^{\mu }\right) =0, \\
\ell _{\mu }^{\mp }\ell _{\pm }^{\mu } &=&\frac{1}{2}\left( x_{\mu }\mp
iy_{\mu }\right) \left( x^{\mu }\pm iy^{\mu }\right) =-1.
\end{eqnarray}%
These new basis vectors are useful since they satisfy the relation, 
\begin{equation}
b^{\mu\nu} = x^{\mu}y^{\nu} - y^{\nu}x^{\mu},  
\end{equation}
which further implies that $\ell _{\nu }^{\pm }$ are the eigenvectors of $b_{\nu }^{\mu }$, with eigenvalues $\pm i$, 
\begin{equation}
b^{\mu \nu }\ell _{\nu }^{\pm }=\left( y^{\mu }x^{\nu }-x^{\mu }y^{\nu
}\right) \frac{1}{\sqrt{2}}\left( x_{\nu }\pm iy_{\nu }\right) =\pm i\ell
_{\pm }^{\mu }.
\end{equation}%
We can then decompose a 4-vector that is orthogonal to both $u^{\mu }$ and $b^{\mu }$ in the following way
\begin{eqnarray}\label{lvec}
A^{\mu}_{\perp} \equiv A_{\perp}^{+}\ell _{+}^{\mu }+A_{\perp}^{-}\ell _{-}^{\mu }, & \quad
A_{\perp}^{\pm } \equiv -\ell _{\mu }^{\mp }A^{\mu }_\perp, 
\end{eqnarray}%
For a traceless, symmetric, second-rank tensor, that is orthogonal to both $%
u^{\mu }$ and $b^{\mu }$, we have that 
\begin{eqnarray}
A_{\indep}^{\mu\nu} \equiv A_{\indep}^{+}\ell _{+}^{\mu }\ell_{+}^{\nu
}+A_{\indep}^{-}\ell_{-}^{\mu }\ell_{-}^{\nu }, & \quad A_{\indep}^{\pm}\equiv\ell_{\mu}^{\mp}\ell _{\nu }^{\mp}A_{\indep}^{\mu \nu }.
\end{eqnarray}
Thus, after implementing this tensor decomposition, $\pi^{\mu\nu}$ will be expressed in terms of 5 scalar independent degrees of freedom, $\pi_\parallel$, $\pi_\perp^{\pm}$, and $\pi_{\indep}^{\pm}$. We shall derive the equations of motion for each of these components. An important observation is that now the '$\pm$' index no longer denotes the particles species but rather our convention for the projections into the subspace orthogonal to the magnetic field and fluid 4-velocity.

\subsubsection{Scalar component}
 We contract \eqref{totalshear} and \eqref{deltashear} with $b^{\mu}b^{\nu}$ and use the tensor decomposition introduced in \eqref{projmag} to obtain the equations of motion for the longitudinal component of the total and relative shear-stress tensor, $\pi_{\parallel}$ and $\delta\pi_{\parallel}$, respectively. The resulting equations are,
\begin{equation}\label{loneq}
\dot{\pi}_{\parallel }+\pi _{\perp }^{\mu}\dot{b}_{\mu}+\Sigma\pi_{\parallel} = \frac{8}{15}\epsilon\sigma _{\parallel }-\frac{4}{3}\pi_{\parallel }\theta -\frac{10}{7}\left( -\frac{1}{2}\pi_{\parallel}\sigma _{\parallel }+\frac{1}{3}\sigma_{\perp }^{\mu }\pi_{\perp \mu}+\frac{1}{3}\pi_{\indep \alpha \beta }\sigma _{\indep}^{\alpha \beta }\right)-\frac{2}{3} \left( \omega^{\mu}_{\perp}\pi_{\perp \mu} + \omega^{\alpha\beta}_{\indep}\pi_{\indep\alpha\beta}\right),
\end{equation}
\begin{equation}\label{delloneq}
\delta\dot{\pi}_{\parallel }+\delta\pi _{\perp }^{\mu}\dot{b}_{\mu}+\Sigma^{'}\delta\pi_{\parallel} = -\frac{4}{3}\delta\pi_{\parallel }\theta -\frac{10}{7}\left( -\frac{1}{2}\delta\pi _{\parallel }\sigma _{\parallel }+\frac{1}{3}\sigma
_{\perp }^{\mu }\delta\pi _{\perp \mu }+\frac{1}{3}\delta\pi_{\indep \alpha \beta }\sigma _{\indep}^{\alpha \beta }\right)-\frac{2}{3} \left( \omega^{\mu}_{\perp}\delta\pi_{\perp \mu} +\omega^{\alpha\beta}_{\indep}\delta\pi_{\indep\alpha\beta}\right).
\end{equation}
We observe that both equations are decoupled, since the term proportional to $B$ vanishes. The only coupling to the magnetic field appears in the term proportional to $\dot{b}^\mu$. Furthermore, due to our assumption of a vanishing chemical potential, the equation of motion for $\delta\pi_{\parallel}$ does not contain a Navier-Stokes-like term, i.e., a term that is proportional to the shear tensor. This implies that $\delta\pi_{\parallel}$ is at least of second-order in an asymptotic gradient expansion and, for this reason, will not contribute in the derivation of a second-order theory of fluid dynamics. We will come back to this point later in this paper. 

\subsubsection{Vector component}
Next, we project Eqs.\ \eqref{totalshear} and \eqref{deltashear} with $b_{\mu}\Xi_{\nu}^{\lambda}$, resulting in an equation of motion for the partially transverse component of the total and relative shear-stress tensor, $\pi_{\perp }^{\nu}$ and $\delta\pi_{\perp }^{\nu}$, respectively,
\begin{equation}
\begin{split}\label{vectormag}
\Xi _{\nu }^{\lambda }\dot{\pi}_{\perp }^{\nu }&+\left(\frac{3}{2}\pi_{\parallel }\Xi^{\lambda \nu}+\pi_{\indep}^{\lambda \nu}\right) \dot{b}_{\nu}+\Sigma\pi_{\perp}^{\lambda }+\frac{B|q|}{5T}
b^{\lambda \nu }\delta\pi_{\perp \nu} \\
=&\frac{8}{15}\epsilon\sigma _{\perp }^{\lambda}-\frac{4}{3}\pi
_{\perp }^{\lambda }\theta +\frac{5}{14}\left( \pi_{\perp }^{\lambda
}\sigma_{\parallel }+\sigma _{\perp }^{\lambda }\pi_{\parallel }\right) -
\frac{5}{7}\left(\sigma_{\perp \nu}\pi_{\indep }^{\lambda \nu}+\pi_{\perp \nu }\sigma_{\indep }^{\lambda \nu }\right) +\frac{1}{2}\pi_{\parallel}\omega^{\lambda}_{\perp}-\pi^{\lambda}_{\indep\nu}\omega^{\nu}_{\perp}-\pi_{\perp\nu}\omega^{\lambda\nu}_{\indep},
\end{split}
\end{equation}
\begin{equation}
\begin{split}\label{delvecmag}
\Xi_{\nu}^{\lambda}\delta\dot{\pi}_{\perp }^{\nu }&+\left(\frac{3}{2}\delta\pi_{\parallel }\Xi ^{\lambda \nu}+\delta\pi_{\indep}^{\lambda \nu }\right) \dot{b}_{\nu}+\Sigma^{'}\delta\pi_{\perp }^{\lambda }+\frac{B|q|}{5T} b^{\lambda \nu }\pi_{\perp \nu} \\
=&-\frac{4}{3}\delta\pi_{\perp }^{\lambda }\theta +\frac{5}{14}\left( \delta\pi _{\perp}^{\lambda}\sigma _{\parallel }+\sigma_{\perp }^{\lambda }\delta\pi _{\parallel }\right) - \frac{5}{7}\left( \sigma_{\perp \nu}\delta\pi _{\indep}^{\lambda \nu}+\delta\pi_{\perp \nu }\sigma_{\indep}^{\lambda \nu }\right) +\frac{1}{2}\delta\pi_{\parallel}\omega^{\lambda}_{\perp}-\delta\pi^{\lambda}_{\indep\nu}\omega^{\nu}_{\perp}-\delta\pi_{\perp\nu}\omega^{\lambda\nu}_{\indep}.
\end{split}
\end{equation}
In this case, the equations for $\pi_{\perp }^{\mu }$ and $\delta{\pi}_{\perp }^{\mu }$ are coupled due to the presence of a magnetic field. We also observe the presence of nonlinear terms that couple the semi-transverse projections of the shear-stress tensor with its scalar and tensor components. 

We see that the term proportional to $\sim b^{\lambda \nu }\pi _{\perp \nu}$ ($\sim b^{\lambda \nu }\delta\pi _{\perp \nu}$) is a first order term that couples different components of $\pi_{\perp}^\mu$ ($\delta\pi_{\perp}^\mu$), which is an inconvenient feature that renders deriving the Navier-Stokes limit of the equations more complicated. We can eliminate this unpleasant feature by further contracting the above equations with $\ell _{\lambda }^{\pm}$, which are the eigenvectors of $b^{\mu \nu}$, leading to linearly independent equations for each semi-transverse components of $\pi_{\perp}^\pm$ ($\delta\pi_{\perp}^\pm$), 
\begin{equation}\label{lvecomp}
\begin{split}
\dot{\pi}_{\perp }^{\mp }&+\pi_{\perp }^{\mp }\ell _{\mp }^{\nu }\dot{
\ell}_{\nu }^{\pm }-\left( \frac{3}{2}\pi _{\parallel }\ell^{\nu}_{\pm}-\pi _{\indep}^{\mp}\ell _{\mp }^{\nu }\right) \dot{b}_{\nu }+\Sigma\pi _{\perp }^{\mp }\mp \frac{iB|q|}{5T}\delta\pi_{\perp}^{\mp } \\
&=\frac{8}{15}\epsilon\sigma _{\perp }^{\mp }-\frac{4}{3}\pi
_{\perp }^{\mp }\theta +\frac{5}{14}\left( \pi _{\perp }^{\mp }\sigma_{\parallel }+\pi _{\parallel }\sigma _{\perp }^{\mp }\right) +\frac{5}{7} \left(\pi_{\perp }^{\pm }\sigma _{\indep}^{\mp}+\pi_{\indep}^{\mp}\sigma _{\perp }^{\pm }\right) +\frac{1}{2}\pi_{\parallel}\omega^{\mp}+ \omega_{\perp}^{\pm}\pi^{\mp}_{\indep} + \omega_{\indep}^{\mp}\pi^{\pm}_{\perp}.
\end{split}\end{equation}
\begin{equation}\label{dellvecomp}\begin{split}
\delta\dot{\pi}_{\perp }^{\mp }&+\delta\pi _{\perp }^{\mp }\ell _{\mp }^{\nu }\dot{%
\ell}_{\nu }^{\pm }-\left( \frac{3}{2}\delta\pi _{\parallel }\ell^{\nu}_{\pm}-\delta\pi_{\indep}^{\mp }\ell_{\mp }^{\nu }\right)\dot{b}_{\nu}+\Sigma^{'}\delta\pi_{\perp}^{\mp}\mp \frac{iB|q|}{5T}\pi_{\perp}^{\mp } \\
&=-\frac{4}{3}\delta\pi_{\perp}^{\mp}\theta +\frac{5}{14}\left(\delta\pi_{\perp }^{\mp}\sigma_{\parallel}+\delta\pi _{\parallel }\sigma_{\perp }^{\mp }\right) +\frac{5}{7} \left( \delta\pi_{\perp }^{\pm }\sigma _{\indep}^{\mp }+\delta\pi_{\indep}^{\mp }\sigma _{\perp }^{\pm}\right) +\frac{1}{2}\delta\pi_{\parallel}\omega^{\mp}+ \omega_{\perp}^{\pm}\delta\pi^{\mp}_{\indep} + \omega_{\indep}^{\mp}\delta\pi^{\pm}_{\perp} .
\end{split}
\end{equation}
Naturally, the resulting equations of motion still display a coupling between $\pi_{\perp }^{\mp }$ and $\delta\pi_{\perp }^{\mp }$, due to the magnetic field.

\subsubsection{Tensor component}
Finally, we project Eqs.~\eqref{totalshear} and \eqref{deltashear} with $\Xi _{\mu \nu }^{\lambda \rho }$ to obtain the equations of motion for the transverse components of the total and relative shear-stress tensor, $\pi_{\indep}^{\alpha \beta}$ and $\delta \pi_{\indep}^{\alpha \beta}$, respectively, 
\begin{equation}
    \begin{split}
\Xi_{\alpha \beta }^{\lambda \rho }&\dot{\pi}_{\indep}^{\alpha \beta
}+2\Xi_{\alpha \beta }^{\lambda \rho }\pi_{\perp }^{\alpha }\dot{b}^{\beta
}+\Sigma\pi_{\indep}^{\lambda \rho }+\frac{B|q|}{5T}
\left(b^{\alpha \lambda }\delta\pi_{\indep \alpha }^{\rho}+b^{\alpha\rho}\delta\pi_{\indep \alpha }^{\lambda }\right) \\
& =  \frac{8}{15}\epsilon\sigma_{\indep}^{\lambda \rho }-\frac{4}{3}\pi_{\indep}^{\lambda \rho }\theta -\frac{5}{7}\left(\pi_{\parallel}\sigma_{\indep}^{\lambda \rho }+\sigma_{\parallel }\pi_{\indep}^{\lambda
\rho }\right)  \\
&-\frac{5}{7}\left[ \pi_{\indep \alpha }^{\lambda }\sigma_{\indep}^{\rho
\alpha }+\pi_{\indep \alpha }^{\rho }\sigma _{\indep}^{\lambda \alpha }-\pi_{\perp }^{\lambda }\sigma_{\perp }^{\rho }-\pi_{\perp }^{\rho }\sigma_{\perp }^{\lambda }+\Xi ^{\lambda \rho }\left( \pi _{\perp \alpha }\sigma_{\perp }^{\alpha }-\pi_{\indep \alpha \beta }\sigma_{\indep}^{\alpha \beta}\right) \right] \\
&-  \pi_{\parallel}\omega_{\indep}^{\lambda\rho} + \pi_{\perp}^{\lambda}\omega_{\perp}^{\rho}+\pi_{\perp}^{\rho}\omega_{\perp}^{\lambda}-\pi_{\indep\alpha}^{\lambda}\omega_{\indep}^{\rho\alpha}-\pi_{\indep\alpha}^{\rho}\omega_{\indep}^{\lambda\alpha}-\Xi ^{\lambda \rho }\left( \pi _{\perp \alpha }\omega_{\perp }^{\alpha }-\pi _{\indep\alpha \beta }\omega_{\indep}^{\alpha \beta}\right)\hspace{4.5em}
\end{split}\end{equation}
\begin{equation}\begin{split}
\Xi_{\alpha \beta }^{\lambda \rho }&\delta\dot{\pi}_{\indep}^{\alpha \beta
}+2\Xi_{\alpha \beta }^{\lambda \rho }\delta\pi_{\perp }^{\alpha }\dot{b}^{\beta
}+\Sigma^{'}\delta\pi_{\indep}^{\lambda \rho }+\frac{B|q|}{5T}
\left(b^{\alpha \lambda}\pi_{\indep\alpha }^{\rho }+b^{\alpha \rho}\pi_{\indep \alpha}^{\lambda }\right)\\
=& -\frac{4}{3}\delta\pi_{\indep}^{\lambda \rho }\theta -\frac{5}{7}\left(\delta\pi_{\parallel}\sigma_{\indep }^{\lambda\rho}+\sigma_{\parallel}\delta\pi_{\indep}^{\lambda\rho }\right) \\
&-\frac{5}{7}\left[\delta\pi_{\indep \alpha }^{\lambda}\sigma_{\indep}^{\rho
\alpha }+\delta\pi_{\indep \alpha }^{\rho }\sigma _{\indep}^{\lambda \alpha }-\delta\pi_{\perp }^{\lambda }\sigma_{\perp }^{\rho }-\delta\pi_{\perp }^{\rho }\sigma_{\perp }^{\lambda }+\Xi^{\lambda \rho }\left( \delta\pi _{\perp \alpha }\sigma_{\perp }^{\alpha }-\delta\pi_{\indep \alpha \beta }\sigma_{\indep}^{\alpha \beta
}\right) \right] \\
& - \delta\pi_{\parallel}\omega_{\indep}^{\lambda\rho} +\delta\pi_{\perp}^{\lambda}\omega_{\perp}^{\rho}+\delta\pi_{\perp}^{\rho}\omega_{\perp}^{\lambda}-\delta\pi_{\perp\alpha}^{\lambda}\omega_{\indep}^{\rho\alpha}-\delta\pi_{\indep\alpha}^{\rho}\omega_{\indep}^{\lambda\alpha}-\Xi^{\lambda \rho }\left(\delta\pi _{\perp \alpha }\omega
_{\perp }^{\alpha }-\delta\pi_{\indep \alpha \beta }\omega_{\indep}^{\alpha \beta}\right)
\end{split}
\end{equation}
Once again, we observe that the equations of motion for $\pi_{\indep}^{\mu\nu}$ and $\delta\pi_{\indep}^{\mu\nu}$ are coupled due to the presence of a magnetic field. Similar to its vector component, the equation of motion for the tensor component also contains nonlinear terms that couples it to the scalar and semi-transverse projections of the shear-stress tensor.

In order to simplify the coupling between different components of $\pi_{\indep}^{\mu\nu}$ ($\delta\pi_{\indep}^{\mu\nu}$) arising from the first-order term proportional to $\sim$ $b^{\alpha \lambda }\pi_{\indep \alpha }^{\rho}$ ($b^{\alpha \lambda}\delta\pi_{\indep \alpha }^{\rho}$), we further project the above equations with $\ell _{\lambda }^{\pm }\ell _{\rho}^{\pm }$. This projection allows us to obtain linearly independent equations for each transverse component $\pi_{\indep}^{\pm}$ ($\delta\pi_{\indep}^{\pm}$), making subsequent calculations simpler. The result is,
\begin{equation}
    \begin{split}\label{tencomp}
        \dot{\pi}_{\indep}^{\mp }+&2\pi _{\indep }^{\mp}\ell _{\mp }^{\beta }%
\dot{\ell}_{\beta }^{\pm }-2\pi _{\perp }^{\mp }\ell _{\beta }^{\pm }\dot{b}%
^{\beta }+\Sigma \pi _{\indep}^{ \mp}\pm i\frac{2B|q|}{5T}\delta \pi_{\indep}^{\mp} \\
=& \frac{8}{15}\epsilon\sigma_{\indep}^{\mp}-\frac{4}{3}\pi_{\indep}^{\mp}\theta - \frac{5}{7}\left( \pi_{\parallel }\sigma
_{\indep}^{\mp}+\sigma _{\parallel }\pi _{\indep}^{\mp }\right) +
\frac{10}{7}\pi_{\perp }^{\mp }\sigma _{\perp }^{\mp}-\pi_{\parallel}\omega_{\indep}^{\mp}+2\pi_{\perp}^{\mp}\omega_{\perp}^{\mp}
\end{split}\end{equation}
\begin{equation}
\begin{split}\label{deltencomp}
\delta \dot{\pi}_{\indep}^{\mp }+ & 2\delta \pi_{\indep}^{\mp }\ell
_{\mp }^{\beta }\dot{\ell}_{\beta }^{\pm }-2\delta \pi_{\perp }^{\mp }\ell
_{\beta}^{\pm }\dot{b}^{\beta }+\Sigma^{'}\delta \pi
_{\indep}^{\mp}\pm i\frac{2B|q|}{5T}\pi _{\indep }^{\mp} \\
=&-\frac{4}{3}\delta \pi_{\indep}^{\mp }\theta -\frac{5}{7}\left(
\delta \pi _{\parallel }\sigma _{\indep}^{\mp}+\sigma _{\parallel
}\delta \pi_{\indep}^{\mp}\right) +\frac{10}{7}\delta \pi _{\perp
}^{\mp }\sigma _{\perp }^{\mp } -\delta\pi_{\parallel}\omega_{\indep}^{\mp}+2\delta\pi_{\perp}^{\mp}\omega_{\perp}^{\mp}
    \end{split}
\end{equation}

The coupling between $\pi_{\indep}^{\mp}$ and $\delta\pi_{\indep}^{\mp}$ persists due to the presence of a magnetic field.

\subsection{Linear Regime}
\label{sub:linear regime}
Up to this point, we have obtained a theory describing the dynamics of the total and relative shear-stress tensor of a system composed of two particle species. Before explaining how we can obtain a closed set of fluid-dynamical equations for the total shear-stress tensor, we discuss some basic features of the coupled equations derived in the previous section. For this purpose, we consider the linearized equations of motion around a static equilibrium state with temperature, $T$, and a constant magnetic field, $B$. In the linear regime, the equations of motion for the semi-transverse total and relative shear-stress tensors are
\begin{align}
 \label{eq: linvector} \dot{\pi}_{\perp }^{\mp}+\Sigma \pi _{\perp }^{\mp
}\mp \frac{iB|q|}{5T}\delta \pi_{\perp }^{\mp} &=\frac{8}{15}\epsilon\sigma_{\perp }^{\mp}, \\ 
 \label{lindelvector}
\delta \dot{\pi}_{\perp }^{\mp }+\Sigma^{\prime}\delta \pi
_{\perp }^{\mp }\mp \frac{iB|q|}{5T}\pi _{\perp }^{\mp } &=0 ,
\end{align}%
while the linearized equations for the transverse components take the following form,
\begin{align}
\label{linearten}
\dot{\pi}_{\indep}^{\mp}+ \Sigma\pi _{\indep}^{\mp}\pm i\frac{2B|q|}{5T}\delta \pi _{\indep}^{\mp} &=\frac{8}{15}\epsilon\sigma_{\indep}^{\mp}, \\
\label{lineardelten}\delta \dot{\pi}_{\indep}^{\mp }+\Sigma^{'} \delta \pi
_{\indep}^{\mp}\pm i\frac{2B|q|}{5T}\pi _{\indep}^{\mp} &=0.
\end{align}
The longitudinal component of the shear-stress tensor will not be discussed here, since it does not couple to the magnetic field and does not exhibit any novel features, when compared to traditional fluid-dynamical formulations. 

Differentiating equation  (\ref{eq: linvector}) with respect to time, and utilizing equation (\ref{lindelvector}), we obtain an equation of motion for the semi-transverse component of the total shear-stress tensor. A similar procedure can also be applied to the transverse components, resulting in a reduction of the equations to the following form,
\begin{align}
\ddot{\pi}_{\perp }^{\mp }+\left(\Sigma+\Sigma^{'}\right)\dot{\pi}_{\perp }^{\mp }+\left(\Sigma\Sigma^{'}+\Omega^2\right)\pi _{\perp }^{\mp } &=\frac{8}{15}\epsilon\Sigma^{'}\sigma _{\perp }^{\mp }+\frac{8}{15}\epsilon\dot{
\sigma}_{\perp }^{\mp }, \\
\ddot{\pi}_{\indep}^{\mp }+\left(\Sigma+\Sigma^{'}\right)\dot{\pi}_{\indep }^{\mp }+\left(\Sigma\Sigma^{'}+4\Omega^2\right)\pi _{\indep}^{\mp} &=\frac{8}{15}\epsilon\Sigma^{'} \sigma _{\indep}^{\mp }+\frac{8}{15}\epsilon
\dot{\sigma}_{\indep}^{\mp }.
\end{align}
Above, we defined the following quantity, 
\begin{equation}
\Omega \equiv \frac{|q|B}{5T},
\end{equation}
which has dimension of frequency. It is evident that these equations are equivalent to the equations of motion of a forced damped harmonic oscillator. The only difference between the equations of motion for the semi-transverse and transverse components is how the magnetic field contributes to the natural frequency of oscillation of the system.  

We now determine the dispersion relation for these linear equations in the homogeneous limit, in which all terms proportional to the shear tensor vanish. In this case, the dispersion relation resulting from the equation of motion for the semi-transverse component is,
\begin{equation}
-\omega ^{2}+i\omega \left(\Sigma+\Sigma^{'}\right)+\left( \Sigma\Sigma^{'}+\Omega^2\right)=0, 
\end{equation}
with solutions,
\begin{equation}
\omega = \frac{i}{2}\left[\Sigma+\Sigma^{'}\pm\sqrt{\left(\Sigma-\Sigma^{'}\right)^2 - 4\Omega^2}\right] .
\end{equation}%
In the limit of a vanishing magnetic field, $B \rightarrow 0$, the solution reduces to $\omega = i \Sigma$ and $\omega = i \Sigma^{'}$ and the system relaxes exponentially to equilibrium, within timescales determined by the inverse of $\Sigma$ and $\Sigma^{'}$ -- as expected of a dilute gas.

In the presence of a finite magnetic field, the dynamics of the system qualitatively changes when, $4\Omega^2 > \left(\Sigma - \Sigma^{'}\right)^2$. In this case, the system no longer solely relaxes exponentially to equilibrium but also displays an oscillatory dynamics. Using the microscopic expressions for $\Sigma$ and $\Sigma^{'}$, the condition for the onset of oscillatory dynamics is,
\begin{equation}
    \Omega  > \frac{\Sigma^{'} - \Sigma}{2} \Longrightarrow \frac{|q|B}{T} >  \hat{n}_0 \sigma_T^{+-}. 
\end{equation}
Here we recall that $\displaystyle{\Sigma =\frac{3\hat{n}_0}{5}\left(\sigma_T^{+-}+\sigma_T\right)} $ and $\displaystyle{\Sigma^{'}=\frac{\hat{n}_0}{5}\left(5\sigma_T^{+-}+3\sigma_T\right)}$. Thus, the value of the total cross-section for inter-species scattering determines if the system will oscillate back to equilibrium or not. For smaller values of the magnetic field, the system will relax exponentially back to equilibrium, but the relaxation time scales will depend on the value of the magnetic field.

The dispersion relation for the transverse components are obtained by changing $\Omega \rightarrow 2\Omega$. Then, we obtain  
\begin{equation}
\omega = \frac{i}{2}\left[\Sigma+\Sigma^{'}\pm\sqrt{\left(\Sigma-\Sigma^{'}\right)^2 - 16\Omega^2}\right] .
\end{equation}%
In this case, these modes will approach equilibrium with oscillations when the magnetic becomes larger than,
\begin{equation}
    \Omega  > \frac{\Sigma^{'} - \Sigma}{4} \Longrightarrow \frac{2|q|B}{T} >  \hat{n}_0 \sigma_T^{+-}. 
\end{equation}
Thus, the onset of oscillatory dynamics occurs for smaller values of magnetic field for the transverse components.

One crucial question is whether or not a typical Israel-Stewart fluid-dynamical theory can capture the basic features of these solutions in the oscillatory limit. We shall investigate this in the following section, where we will derive second-order fluid dynamics using the order of magnitude approach \cite{Denicol_Rischke}. Furthermore, we shall later discuss if this oscillatory dynamics survives for a rapidly expanding fluid, with a dynamics analogous to that of the quark-gluon plasma produced in ultrarelativistic heavy ion collisions.

\subsection{Truncation scheme}
\label{sub:Truncation scheme}

In this subsection we will derive a second-order fluid-dynamical theory from the equations of motion obtained in the previous subsections. The main idea is to estimate the magnitude of each term in the equations of motion for the relative and total shear-stress tensor using the leading term in an asymptotic gradient expansion solution \cite{rocha2023theories}. We shall demonstrate that, up to second-order in this power-counting scheme, it is possible to re-express the relative shear-stress tensor solely in terms of the total shear-stress tensor and its derivatives.

We start by analyzing the leading term in an asymptotic gradient expansion of solutions of Eqs.\ (\ref{dellvecomp}) and (\ref{deltencomp}). To first order in gradients, we have that the relative components of the shear-stress tensor can be expressed as
\begin{eqnarray}
\delta\pi^{\mp}_{\perp} &=& \pm i\varphi \pi^{\mp}_{\perp} + \mathcal{O}(2),\\
\delta\pi^{\mp}_{\indep} &=& \pm 2 i \varphi \pi^{\mp}_{\indep} + \mathcal{O}(2),
\end{eqnarray}
where $\mathcal{O}(2)$ denotes terms that are of second-order or higher in powers of gradients or in powers of the dissipative currents. We further defined a new variable $\varphi \equiv B|q|/(5T\Sigma^{'})$ for the sake of brevity. We then iterate these first-order solutions back into Eqs.\ (\ref{dellvecomp}) and (\ref{deltencomp}) and re-express all the remaining terms of the equations up to third order. In this case, we obtain that the semi-transverse components of the relative shear-stress tensor can be approximated as,
\begin{equation}\label{appdel}
    \begin{split}
        \Sigma^{'}\delta\pi^{\mp}_{\perp}  = & \ \mp i \varphi\pi^{\mp}_\perp\left(\frac{4\theta}{3} - \frac{5}{14}\sigma_{\parallel}\right) \mp \frac{5i}{7}\varphi\left(\pi^{\pm}_\perp\sigma_{\perp}^{\mp} + 2\pi^{\mp}_{\indep}\sigma^{\pm}_\perp\right)\mp i \varphi \pi^{\mp}_\perp\ell^{\nu}_{\mp}\dot{\ell}^{\pm}_{\nu} \pm 2i\varphi \pi^{\mp}_{\indep} \ell^{\nu}_{\mp}\dot{b}_{\nu}\\ & \ \pm i \varphi \Sigma^{'}\pi^{\mp}_\perp \mp i \varphi \dot{\pi}^{\mp}_\perp  \mp i\varphi \left( 2\omega^{\pm}_{\perp}\pi^{\mp}_{\indep} + \omega^{\mp}_{\indep}\pi^{\pm}_{\perp}\right) + \mathcal{O}(3).
        \end{split}
\end{equation}
While the fully transverse components of the relative shear-stress tensor can be approximated as,
\begin{equation}
    \begin{split}
        \Sigma^{'}\delta \pi _{\indep}^{\mp} = & 
\pm \frac{4}{3}2i\varphi\pi_{\indep}^{\mp
}\theta \pm \frac{5}{7}2i\varphi\pi_{\indep}^{\mp
}\sigma_{\parallel}\pm \frac{10}{7}i\varphi\pi_{\perp}^{\mp }\sigma_{\perp }^{\mp } \mp 2i\Omega\pi_{\indep}^{\mp}\pm 2i\varphi \dot{\pi}_{\indep}^{\mp }\pm 2i\pi_{\indep}^{\mp}\dot{\varphi} \\ &  \pm 4i\varphi\pi_{\indep}^{\mp}\ell_{\mp}^{\beta}\dot{\ell}_{\beta }^{\pm }\pm 2i\varphi\pi_{\perp }^{\mp }\ell_{\beta}^{\pm}\dot{b}^{\beta} \pm 2i\varphi \pi_{\perp}^{\mp}\omega^{\mp}_{\perp} + \mathcal{O}(3).
    \end{split}
\end{equation}
Similarly, $\mathcal{O}(3)$ denotes terms that are of third-order or higher in powers of gradients or in powers of the dissipative currents. 

A closed set of second-order equations of motion for each component of the total shear-stress tensor can then be obtained by substituting the results above into Eqs.\ \eqref{lvecomp} and \eqref{tencomp}, and disregarding all terms of $\mathcal{O}(3)$. Then, the full second-order equations for the semi-transverse shear-stress tensor are,
 \begin{equation}
     \begin{split}\label{finvec}
         \left( 1-\varphi^2\right)\dot{\pi}^{\mp}_{\perp} + \left(\Sigma+\varphi^2 \Sigma^{'}\right)\pi^{\mp}_{\perp} & = \frac{8}{15}\epsilon \sigma^{\mp}_\perp - \left[\left(1-\varphi^2\right)\left(\ell^{\nu}_{\mp}\dot{\ell}^{\pm}_{\nu}+\frac{4\theta}{3} - \frac{5}{14}\sigma_{\parallel}\right)  + \varphi\dot{\varphi}\right]\pi^{\mp}_{\perp}   +\left(1+\varphi^2\right) \left[ \omega^{\mp}_{\indep}+ \frac{5}{7}\sigma^{\mp}_{\indep}\right]\pi^{\pm}_{\perp}\\&+\left(1+2\varphi^2\right) \left[ -\ell^{\nu}_{\mp}\dot{b}_{\nu}+\omega^{\pm}_{\perp}+\frac{5}{7}\sigma^{\pm}_{\perp}\right]\pi^{\mp}_{\indep} +\left(\frac{3}{2}\ell^{\nu}_{\pm}\dot{b}_{\nu}  + \frac{\omega_{\perp}^{\mp}}{2}+\frac{5}{14}\sigma^{\mp}_{\perp}\right)\pi_{\parallel}. 
         \end{split}
 \end{equation}
While the full second-order equations for the fully-transverse shear-stress tensor are,
\begin{equation}
\begin{split}\label{finten}
 \left(1-4\varphi^{2}\right) \dot{\pi}_{\indep }^{\mp} + \left(\Sigma +4\Sigma^{'}\varphi^2 \right)\pi_{\indep}^{\mp}
& =  \frac{8}{15}\epsilon\sigma _{\indep}^{\mp } -\left[ \left(
1-4\varphi ^{2}\right) \left(2\ell_{\mp}^{\beta }\dot{\ell}
_{\beta}^{\pm} + \frac{4}{3} \theta + \frac{5}{7} \sigma
_{\parallel} \right) -4 \varphi \dot{\varphi}\right] \pi_{\indep}^{\mp} -\left( \frac{5}{7}\sigma _{\indep}^{\mp }+\omega_{\indep}^{\mp}\right) \pi
_{\parallel }\\ & + \left( 1+2\varphi ^{2}\right) \left( 2\ell
_{\beta}^{\pm}\dot{b}^{\beta} +\frac{10}{7} \sigma_{\perp }^{\mp}+ 2
\omega_{\perp}^{\mp}\right)\pi_{\perp }^{\mp}.
\end{split}
\end{equation}

We note that the equations for the longitudinal components of the total shear-stress tensor were already independent from the relative longitudinal shear-stress tensor and, thus, did not have to be simplified using any power-counting scheme. Another novel feature of this theory is that the semi-transverse and transverse components display different relaxation times, i.e., they have their own dynamical equations of motion that cannot be trivially recombined into an unique equation of motion for the complete shear-stress tensor.

A shear viscosity can be identified for the longitudinal, semi-transverse and transverse components of the shear-stress tensor. They are,
\begin{equation}\label{shear_viscosities}
    \eta_\parallel = \frac{4 \epsilon}{15 \Sigma} =\eta ,\qquad
    \eta_\perp  = \frac{4 \epsilon}{15\left(\Sigma + \varphi^2 \Sigma^{'}\right)},\qquad
    \eta_{\indep}  =  \frac{4 \epsilon}{15 \left(\Sigma + 4\varphi^2 \Sigma^{'}\right)}.
\end{equation}
These coefficients are positive definite, with $\eta_\perp$ and $\eta_{\indep}$ displaying a significant dependence on the magnetic field. Similar to Ref.~\cite{denicol2018nonresistive}, the longitudinal shear viscosity displays no dependence on the magnetic field and has the same value of the shear viscosity in the absence of a magnetic field. We recall that, 
\begin{equation}
    \varphi = \frac{B|q|}{5T\Sigma^{'}} ,\qquad
     \Sigma =\frac{3\hat{n}_0}{5}\left(\sigma_T^{+-}+\sigma_T\right),\qquad \Sigma^{'}=\frac{\hat{n}_0}{5}\left(5\sigma_T^{+-}+3\sigma_T\right).
\end{equation}

\begin{figure}[H]
    \centering
     \subfigure[\quad  $\tau_{\perp}$ assuming negative values for $\varphi(\Omega) > 0.5$. ]{\includegraphics[width=0.4\linewidth]{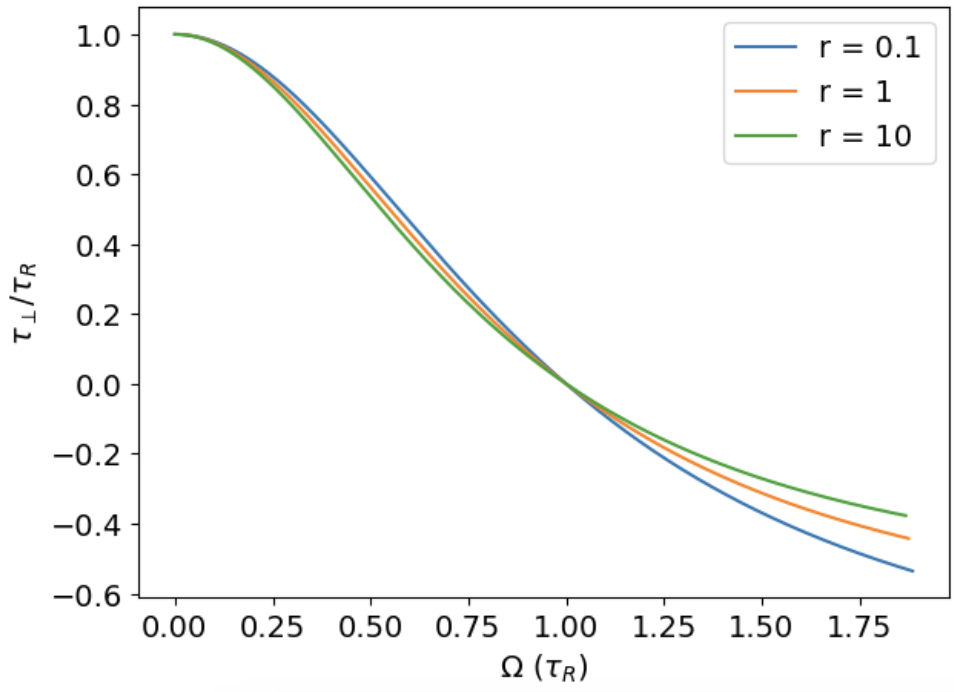}}
     \quad
     \subfigure[\ \ $ \tau_{\indep}$ assuming negative values for $\varphi(\Omega) > 1$. ]{\includegraphics[width=0.41\linewidth]{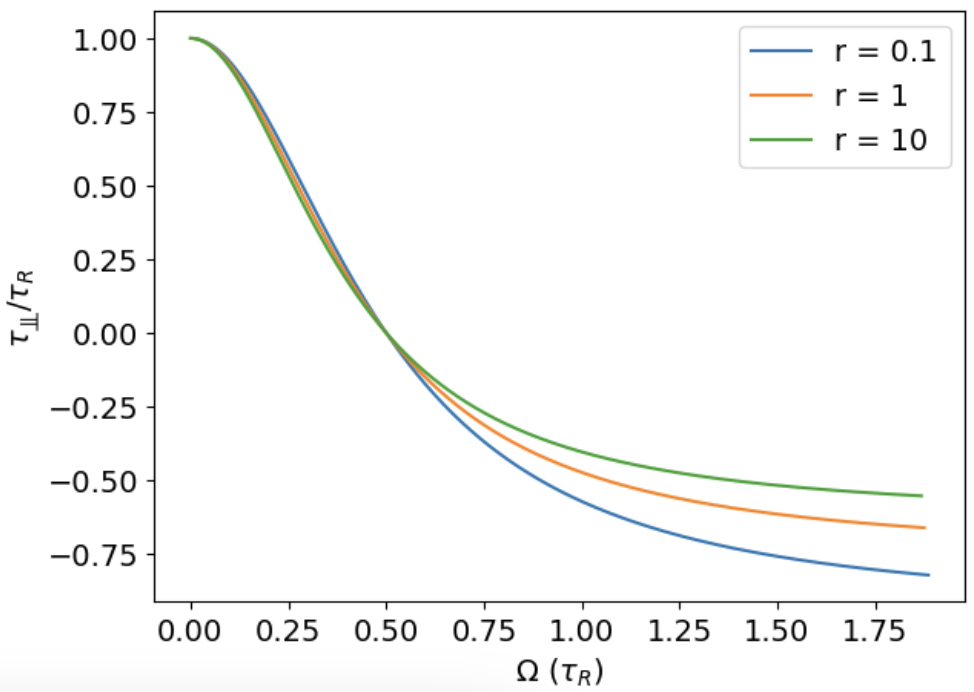}}
    \caption{Relaxation times for the a) semi-transverse and b) transverse components of the shear-stress tensor as a function of $\Omega$, for different values of $\displaystyle{ r = \sigma^{+-}_T/\sigma^{++}_T}$.}
    \label{fig: figtau}
\end{figure}

Similarly, a relaxation time for the longitudinal, semi-transverse and transverse components can be respectively identified as,
\begin{equation}\label{relaxation times expression}
    \tau_\parallel = \frac{1}{\Sigma} =\tau_\pi ,\qquad
    \tau_\perp  = \frac{1-\varphi^2}{\Sigma + \varphi^2 \Sigma^{'}},\qquad
    \tau_{\indep}  =  \frac{1-4\varphi^2}{\Sigma + 4\varphi^2 \Sigma^{'}}.
\end{equation}
Note that, in the limit of vanishing magnetic field, $\varphi \longrightarrow 0$, all the relaxation times become identical and match the usual relaxation time derived for Israel-Stewart theory, $\tau_\pi$. 
Nevertheless, in the limit of moderately large magnetic fields, these relaxation times can differ significantly. In particular, we can see that if, $4\varphi^2 > 1$, the relaxation time of the transverse component of the shear-stress tensor will even become negative. Further increasing $\varphi$, such that $\varphi^2 >1$, will also render the relaxation time for the semi-transverse component negative.  This behaviour is expressed in Fig.~\ref{fig: figtau}, where we denote the ratio between the cross sections as, $r \equiv \sigma^{+-}_T/\sigma^{++}_T$. Negative relaxation times are unphysical and lead to linear instabilities of the global equilibrium state. This indicates that the truncated theory derived in this section breaks down for such values of magnetic fields (or for such values of $\varphi$). We note that $\varphi$ is proportional to $\Omega \tau_\pi$ and, thus, if this quantity is larger than one, it indicates that the frequency of oscillation emerging due to the magnetic field is of the same order as the inverse (longitudinal) relaxation time. In other words, this implies that the period of oscillation is not large relative to the relaxation time scale and oscillation phenomena cannot be neglected.

In the following section, we shall investigate this physics in a simple dynamical model, Bjorken flow \cite{Bjorken}. In this case, we shall verify that oscillations do indeed become important when the frequency of oscillations $\Omega$ becomes large with respect to the inverse relaxation time scale. Thus, the breakdown of this Israel-Stewart-like theory for each component of the shear-stress tensor does occur when oscillatory phenomena emerge. 

\section{Bjorken Flow}
\label{sec:Bjorken flow}
So far, we have obtained the equations of motion for the 5 components of the shear-stress tensor, decomposed with respect to the direction of the magnetic field. In our simplified kinetic description, we found that three of these components evolve independently. Furthermore, we have identified that, in the presence of relatively large magnetic fields, the relaxation times appearing in the equations of motion for the transverse components of the shear-stress tensor become negative -- a clearly unphysical feature. We further argued that this unphysical scenario may be connected to the emergence of oscillatory dynamics for the shear-stress tensor that cannot be captured by the Israel-Stewart-like equations derived so far. 
Our next step is to investigate the emergence of such oscillatory behavior using a simplified solution for expanding plasmas: Bjorken flow \cite{Bjorken}.

Bjorken flow is a toy model for the longitudinal fluid-dynamical expansion that takes place in ultrarelativistic heavy-ion collisions. It describes a boost-invariant, longitudinally (with respect to the beam direction) expanding medium. The system is also traditionally assumed to be isotropic and homogeneous in the transverse place (relative to the beam axis) --  we note that we will break the first assumption by introducing a magnetic field in the transverse plane.  

Thus, Bjorken flow is a highly symmetric flow configuration, making it possible to solve the equations of motion for the shear-stress tensor with simple numerical schemes and gain insights into the theory beyond just linear approximations. In this scenario, the spacetime is conveniently described using hyperbolic coordinates, $\tau$, $\xi$, $x$ and $y$, 
\begin{equation}
\tau = \sqrt{t^2-z^2}, \hspace{.5cm} \xi = \frac{1}{2} \mathrm{ln} \left( \frac{t+z}{t-z} \right).
\end{equation}
where $\tau$ is the proper time, $\xi$ is the spacetime rapidity, with $(t,x,y,z)$ being the usual Cartesian coordinates. In this coordinate system, the metric tensor is given by
\begin{equation}
g_{\mu\nu} =  \mathrm{diag} \, (g_{\tau\tau}, g_{xx}, g_{yy}, g_{\xi \xi}) = \mathrm{diag} \, (1, -1, -1, -\tau^2),
\end{equation}
with the only non-zero Christoffel symbols being
\begin{equation}
\Gamma^\tau_{\xi \xi} = \tau, \hspace{.3cm} \Gamma^{\xi}_{\tau \xi} = \Gamma^{\xi}_{\xi \tau} = \frac{1}{\tau}.
\end{equation}
Naturally, in this coordinate system all space-time derivatives appearing in the equations of motion must be replaced by covariant derivatives, i.e., $\partial_\mu \rightarrow D_\mu $. 

In Bjorken flow, the fluid 4-velocity is static $u^\mu=(1,0,0,0)$ and our basis elements can be expressed as:
\begin{align*}
    x^{\mu} &= (0,0,1,0),\\
    y^{\mu} &= (0,0,0,\frac{1}{\tau}),\\
    b^{\mu} &= (0,1,0,0),
   %l^{\mu}_{\pm} &=\frac{1}{\sqrt{2}} (0,0,1,\pm i) 
\end{align*}
where the magnetic field, $b^\mu$, was chosen to be in the transverse direction relative to the beam axis. The shear tensor in Bjorken flow is given by the covariant derivatives of the 4-velocity and can be calculated to be \cite{Denicol_Rischke}, 
\begin{equation}
    \sigma_{\mu\nu} = \Delta_{\mu\nu}^{\alpha \beta}D_\alpha u_\beta=\textrm{diag} \left(0,\frac{1}{3\tau},\frac{1}{3\tau},-\frac{2\tau}{3} \right).
\end{equation}
Its different components with respect to the direction of the magnetic field are,
\begin{subequations}
\begin{align}
    \sigma_{\parallel}&= b_{\mu}b_{\nu}\sigma^{\mu\nu} = \frac{1}{3\tau},\\
    \sigma^{\pm}_{\perp} & = \ell ^{\mp}_{\mu} b_{\nu}\sigma^{\mu\nu} = 0,\\
    \sigma^{\pm}_{\indep} & = \ell^{\pm}_{\mu}\ell^{\pm}_{\nu} \sigma^{\mu\nu}_{\indep} = \ell^{\pm}_{\mu}\ell^{\pm}_{\nu} \sigma^{\mu\nu} = \frac{1}{2\tau}.
    \end{align}
    \end{subequations}
Finally, the expansion rate is given by, $\displaystyle{\theta = D_\mu u^\mu = 1/\tau}$. Further, the shear-stress tensor in Bjorken flow is expressed as
\begin{equation}
      \pi^{\mu\nu} = \pi_{\parallel}b^\mu b^\nu + \left(\frac{\pi^{++}_{\indep} + \pi^{--}_{\indep} - \pi_\parallel }{2} \right) x^\mu x^\nu - \left(\frac{\pi^{++}_{\indep} + \pi^{--}_{\indep} + \pi_\parallel }{2}\right) y^\mu y^\nu, 
      \label{oops}
\end{equation}
with the magnetic field pointing in the x--direction breaking the degeneracy between the shear-stress tensor components in the transverse plane relative to the beam axis \footnote{In the absence of a magnetic field in the transverse plane, the shear-stress tensor in Bjorken flow has the following general form $\pi^{\mu\nu}$ = \textrm{diag}(0,$\pi$/2,$\pi$/2,-$\pi/\tau^2$) }. Given that $\sigma^{\pm}_\perp = 0$, we can remove this component of the shear-stress tensor by setting its initial value to zero -- this is why this component is not included in the decomposition \eqref{oops}. Finally, we consider the following equation of state (for 2 types of particles having 3 quarks each with two spins),
\begin{equation}
    \epsilon  = \frac{3\times 2\times 2\times  3}{\pi^2}T^4.
\end{equation}
 
 Using the results outlined above, the relevant fluid-dynamical equations reduce to, 
\begin{align}
      \frac{d \epsilon}{d\tau}  & 
      = \frac{\pi_\parallel}{2\tau} + \frac{\pi_{\indep}}{2\tau}-\frac{4\epsilon}{3\tau},\label{energy1}\\
      \frac{d}{d\tau}\left(\frac{\pi_\parallel}{\epsilon}\right) + \frac{1}{\tau_\pi}\frac{\pi_\parallel}{\epsilon} & = \frac{8}{45\tau}+\frac{5}{21\tau}\frac{\pi_\parallel}{\epsilon}-\frac{5}{21\tau}\frac{\pi_{\indep}}{\epsilon} -\frac{\pi_\parallel}{\epsilon^2}\left( \frac{\pi_\parallel + \pi_{\indep}}{2\tau}\right),\label{par1}\\
     \frac{d}{d\tau}\left(\frac{\pi_{{\indep}}}{\epsilon}\right) + \frac{1}{\tau_{\indep}}\frac{\pi_{\indep}}{\epsilon} & = \frac{1}{1-4\varphi^2} \frac{8}{15\tau} - \frac{5}{21\tau}\frac{\pi_{\indep}}{\epsilon} - \frac{1}{1-4\varphi^2}\frac{5}{7\tau}\frac{\pi_\parallel}{\epsilon}+\frac{4\varphi\dot{\varphi}}{1-4\varphi^2}\frac{\pi_{\indep}}{\epsilon} - \frac{\pi_{\indep}}{\epsilon^2}\left(\frac{\pi_\parallel + \pi_{\indep}}{2\tau}\right),\label{secondor}
 \end{align}
 where we defined the variable $\pi_{\indep} \equiv \pi^{-}_{\indep} + \pi^{+}_{\indep}$.
 It is useful to note that, \begin{equation*}
\dot{\varphi} = -\varphi \left( \frac{1}{\tau} + 2\frac{\dot{T}}{T}\right), \qquad
\frac{\dot{T}}{T}  = \frac{1}{4\tau}\left(\frac{\pi_\parallel+\pi_{\indep}}{2\epsilon} -\frac{4}{3}\right).
\end{equation*}
Equation \eqref{energy1} corresponds to the continuity equation related to energy conservation, while Eqs.~\eqref{par1}-\eqref{secondor} correspond to Eqs.~\eqref{loneq} and \eqref{finten}, respectively. We will solve equations \eqref{energy1}--\eqref{secondor}, considering a choice of cross section that satisfies $\Sigma^{'} = 4 \Sigma/3$, i.e., we considered $r$ $ \equiv\sigma^{+-}_T/\sigma_T$ = 1. We further assume that the system is initially at equilibrium, at an initial time of $\tau_0=0.1$ fm and an initial energy density of $\epsilon_0(\tau_0) = 1000$ $\textrm{fm}^{-4}$. The equation of motion for the magnetic field is taken from Maxwell's equations \cite{denicol2019resistive}, 
\begin{equation}\label{Bfield}
    \dot{B} + B\theta = 0\quad \Longrightarrow \quad B \sim \left(\frac{\tau_0}{\tau}\right).
\end{equation}
We shall consider simulations with various initial values of the magnetic field. 

\begin{figure}[H]
    \centering
    \subfigure[\ Longitudinal component. ]{\includegraphics[width=0.4\linewidth]{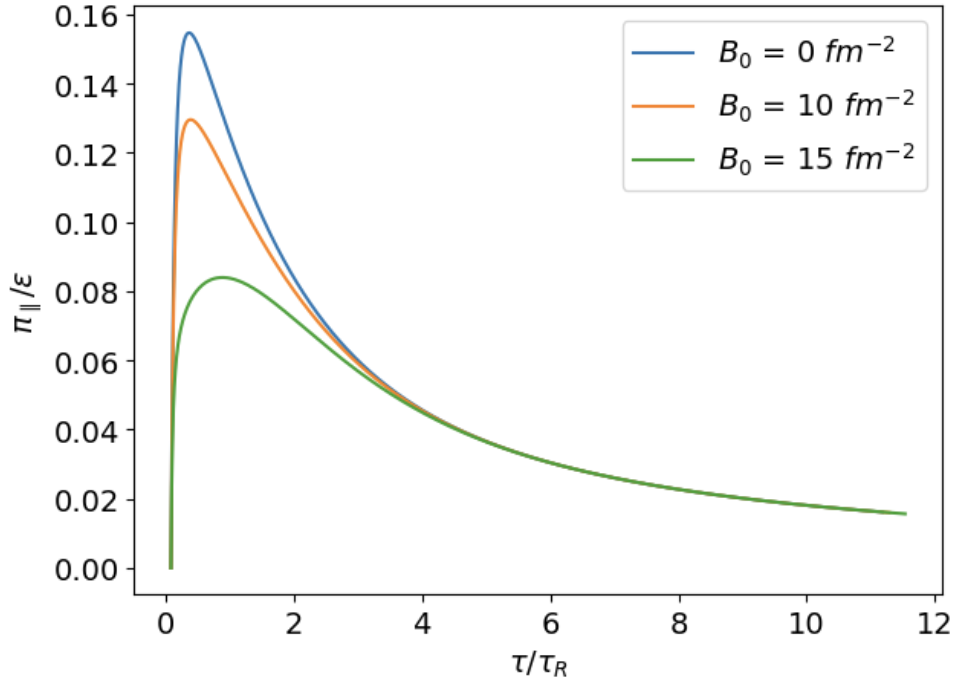}}
    \quad
    \subfigure[\ Transverse component. ]{\includegraphics[width=0.4\linewidth]{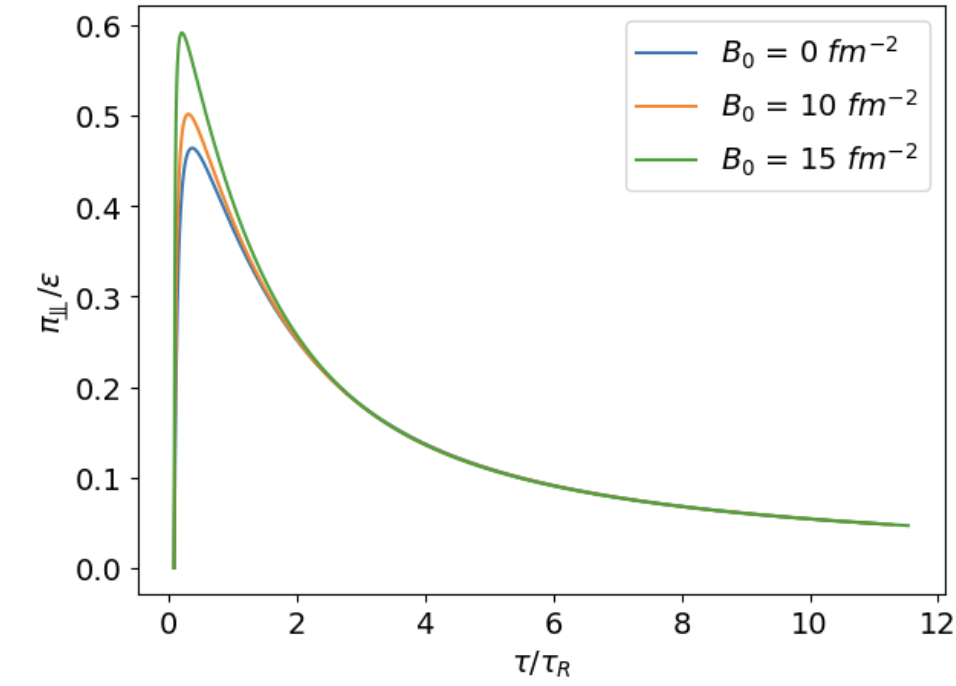}}
    \caption{Time evolution of the a) longitudinal and b) transverse components of the shear-stress tensor for several values of the initial magnetic field, $B_0$. All simulations were performed for $\eta/s$ = 1.}
    \label{fig: figbr.}
\end{figure}

We fix the value of $\Sigma$ so that the shear viscosity to entropy density ratio in the absence of a magnetic is given by $\eta/s$ = 1, cf.~Eq.~\eqref{shear}. Our numerical solutions for the longitudinal (left panel) and transverse (right panel) components of the shear-stress tensor are depicted in Fig.~\ref{fig: figbr.}. For the value of shear viscosity considered, we observe that the Israel-Stewart-like theory derived to describe each of these components is only effective in the region where the magnetic field is smaller than $B_0 \sim 15$ fm$^{-2}$ -- otherwise the transverse relaxation time becomes negative. For larger values of shear viscosity, the relaxation time becomes negative for even smaller values of the magnetic field. This is an extreme restriction on the applicability of this theory. Nevertheless, in Fig.~\ref{fig: figbr.} we see a moderate effect of the magnetic on the magnitude of the shear-stress tensor at the early stages of the evolution, $\tau \sim \tau_R$. When $\tau \gg \tau_R$, the effect of the magnetic field completely disappears.

Hence, to attain an accurate depiction of the system under larger magnetic fields, it is necessary to go back to the more fundamental coupled equations [Eqs. \eqref{loneq} - \eqref{delloneq}, %\eqref{lvecomp} - \eqref{dellvecomp}, 
\eqref{tencomp} - \eqref{deltencomp}], where no truncation scheme has been imposed. In principle, these equations still capture the oscillatory dynamics that we anticipate but are absent when employing truncated equations. The above mentioned set of equations takes following form in Bjorken flow:

 \begin{align}
      \frac{d \epsilon}{d\tau}  & 
      = \frac{\pi_\parallel}{2\tau} + \frac{\pi_{\indep}}{2\tau}-\frac{4\epsilon}{3\tau},\label{energy}\\
      \frac{d}{d\tau}\left(\frac{\pi_\parallel}{\epsilon}\right) + \Sigma\frac{\pi_\parallel}{\epsilon}& = \frac{8}{45\tau}+\frac{5}{21\tau}\frac{\pi_\parallel}{\epsilon}-\frac{5}{21\tau}\frac{\pi_{\indep}}{\epsilon} -\frac{\pi_\parallel}{\epsilon^2}\left( \frac{\pi_\parallel + \pi_{\indep}}{2\tau}\right),\label{par}\\
      \frac{d}{d\tau}\left(\frac{\pi_{\indep}}{\epsilon}\right) + \Sigma\frac{\pi_{\indep}}{\epsilon} - \frac{2|q|B}{5T} \frac{\delta\hat{\pi}_{\indep}}{\epsilon}& = \frac{8}{15\tau}-\frac{5}{7\tau}\frac{\pi_\parallel}{\epsilon}-\frac{5}{21\tau}\frac{\pi_{\indep}}{\epsilon}-\frac{\pi_{\indep}}{\epsilon^2}\left( \frac{\pi_\parallel + \pi_{\indep}}{2\tau}\right),\label{perp}\\
      \frac{d}{d\tau}\left(\frac{\delta\hat{\pi}_{\indep}}{\epsilon}\right) + \Sigma^{'} \frac{\delta\hat{\pi}_{\indep}}{\epsilon} + \frac{2|q|B}{5T} \frac{\pi_{\indep}}{\epsilon}& = -\frac{5}{21\tau}\frac{\delta\hat{\pi}_{\indep}}{\epsilon}-\frac{\delta\hat{\pi}_{\indep}}{\epsilon^2}\left( \frac{\pi_\parallel + \pi_{\indep}}{2\tau}\right),\label{relperp}
 \end{align}
where we defined the following variables $\pi_{\indep} \equiv \pi^{-}_{\indep} + \pi^{+}_{\indep}$,
$\delta\pi_{\indep} \equiv \delta\pi^{-}_{\indep} - \delta\pi^{+}_{\indep}$, and $\delta\pi_{\indep} = i\delta\hat{\pi}_{\indep}$. We again consider $\Sigma^{'} $ = $4\Sigma/3$.

Our objective is to solve \eqref{energy}--\eqref{relperp} to investigate the behavior of the longitudinal and transverse components of the shear-stress tensor. Maxwell's equations remain unaltered, and consequently, the evolution of $B$ follows from Eq \eqref{Bfield}. As before, we assume that the system is at equilibrium at an initial time of $\tau_0=0.1$ fm, with an initial energy density of $\epsilon_0(\tau_0)=1000$ fm$^{-4}$. This time, we enforce the presence of stronger magnetic fields, while also considering different $\eta/s$ values.

\begin{figure}[H]
    \centering
    \subfigure[\ Longitudinal component. ]{\includegraphics[width=0.4\linewidth]{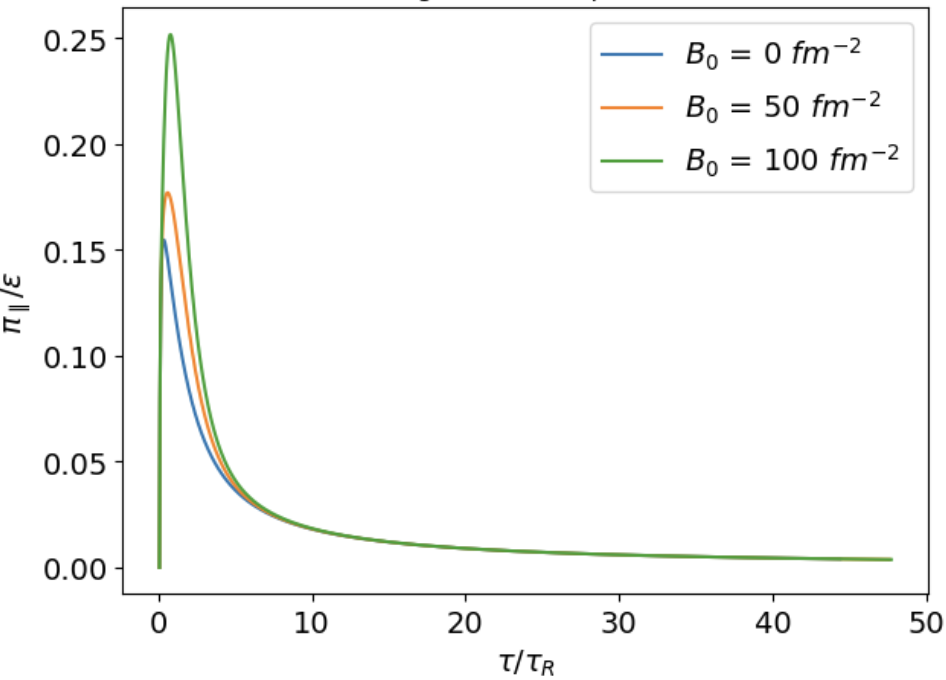}}
    \quad
    \subfigure[\ Transverse component. ]{\includegraphics[width=0.4\linewidth]{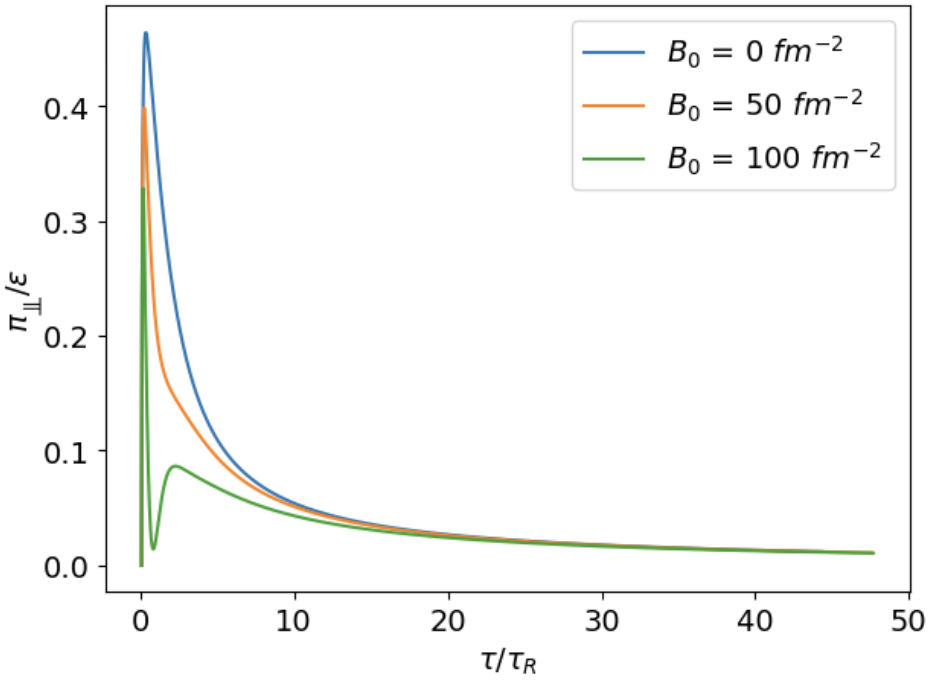}}
    \caption{Time evolution of the a) longitudinal and b) transverse components of the shear-stress tensor for several values of the initial magnetic field, $B_0$. All simulations were performed for $\eta/s$ = 1.}
    \label{fig: figbr1}
\end{figure}
 
\begin{figure}[H]
    \centering
    \subfigure[\ Longitudinal component]{\includegraphics[width=0.4\linewidth]{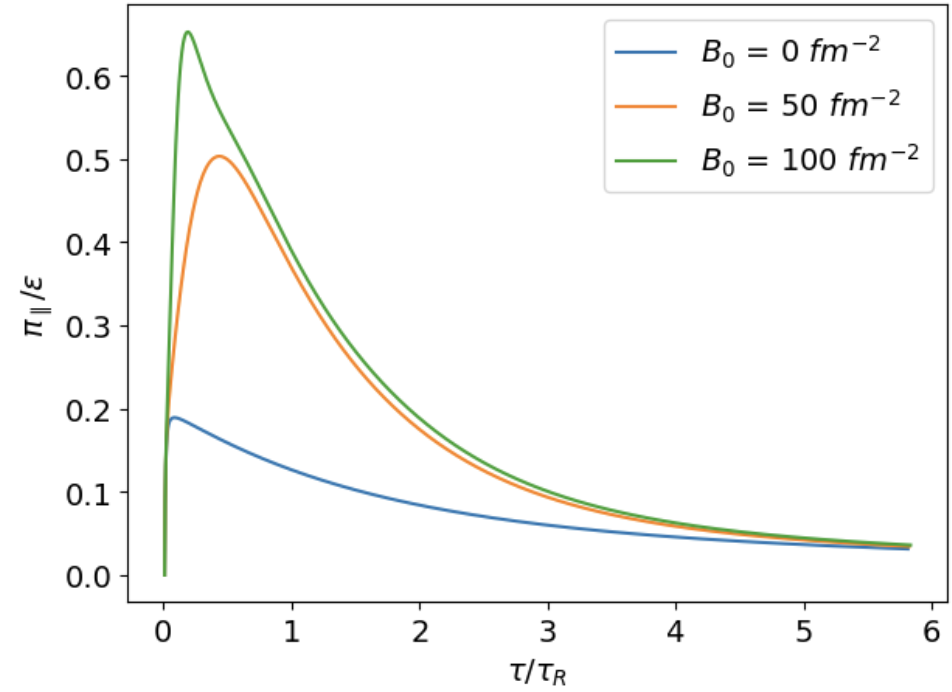}}
    \quad
    \subfigure[\ Transverse component ]{\includegraphics[width=0.4\linewidth]{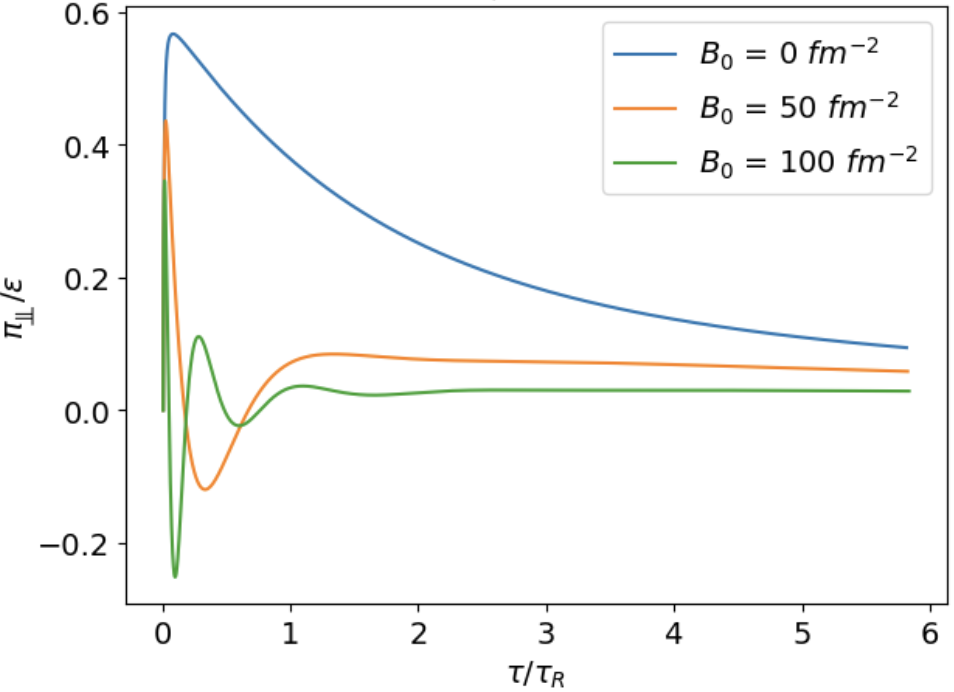}}
    \caption{Time evolution of the a) longitudinal and b) transverse components of the shear-stress tensor for several values of the initial magnetic field, $B_0$. All simulations were performed for $\eta/s$ = 10.} 
    \label{fig: figbr2}
\end{figure}
In Fig.~\ref{fig: figbr1} we show our results for the longitudinal (left panel) and transverse (right panel) components of the shear-stress tensor for $\eta/s=1$ and an initial magnetic field of $B_0=0$, $50$, $100$ fm$^{-2}$. In \ref{fig: figbr2}, we show the same quantities for a larger value of shear viscosity $\eta/s=10$. Indeed, we do observe the appearance of oscillatory dynamics for the transverse component of the shear-stress tensor -- it no longer simply relaxes exponentially to zero. As expected, this oscillatory behavior only emerges for larger values of the magnetic field, that could not be probed in our previous simulations, for the truncated second-order equations. We note that as the value of $\eta/s$ increases, the oscillations become more pronounced. This phenomenon can be attributed to the faster changes in the source term, i.e. the respective components of $\sigma^{\mu\nu}$, when $\eta/s$ is smaller. Larger $\eta/s$ values correspond to longer relaxation times, allowing the magnetic field sufficient time to induce prominent oscillations in the system. On the other hand, with smaller $\eta/s$ values, relaxation times are shorter and oscillations do not have sufficient time to develop.

\begin{figure}[H]
    \centering
    \subfigure[\ $\eta/s = 1$]{\includegraphics[width=0.4\linewidth]{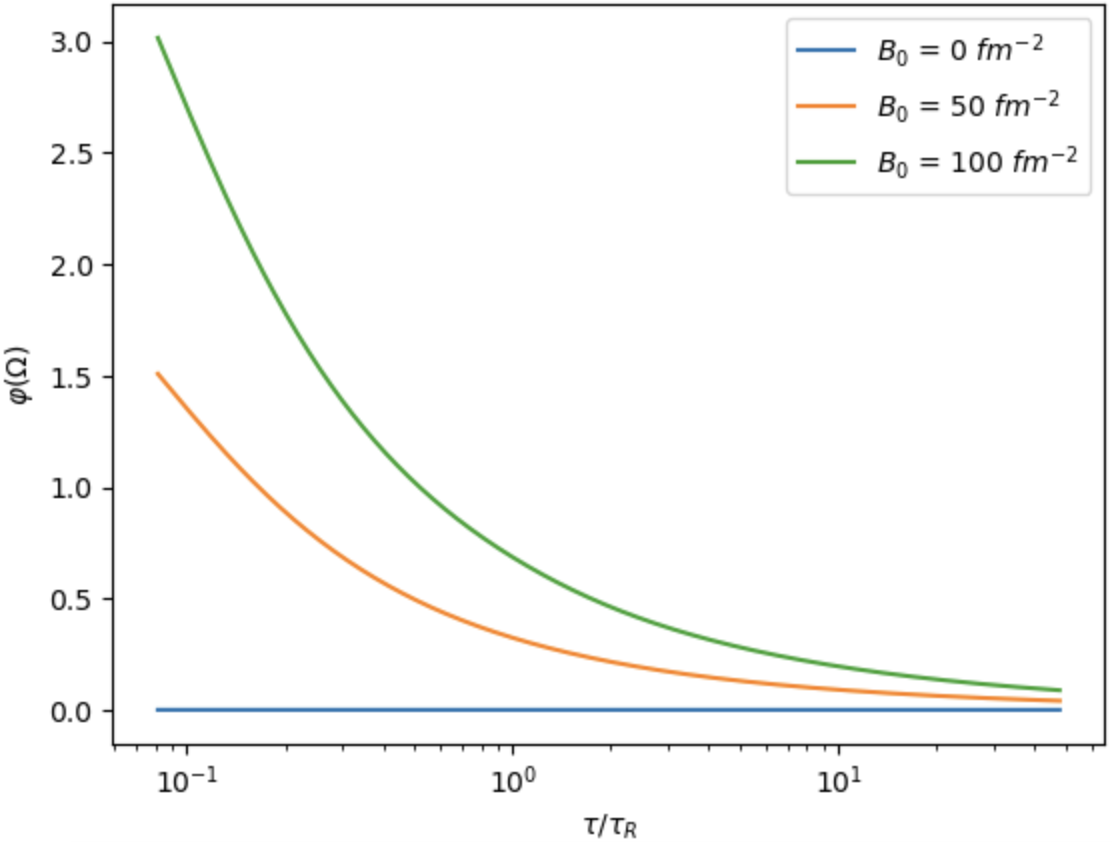}}
    \quad
    \subfigure[\ $\eta/s = 10$ ]{\includegraphics[width=0.4\linewidth]{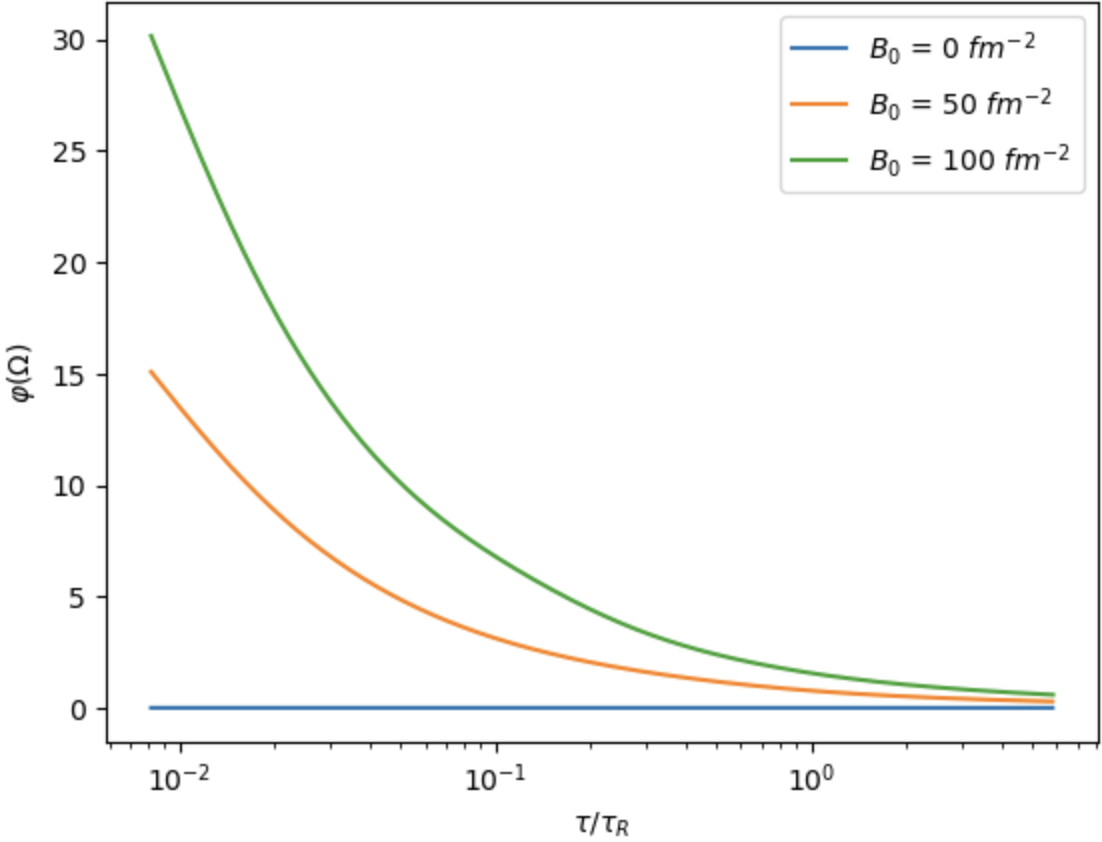}}
    \caption{The dimensionless variable $\varphi(\Omega)$ and a function of $\tau/\tau_R$ for different $\eta/s=1$ (left panel) and $\eta/s=10$ (right panel). }
    \label{fig: figphi}
\end{figure}

The oscillatory dynamics can also be better comprehended by analysing the magnitude of the parameter $\varphi$ in our simulations. For the transverse component, we expect that oscillatory dynamics occurs when the transverse relaxation time is negative, which happens when $\varphi > 0.5$. In Fig.~\ref{fig: figphi} we show $\varphi$ as a function of $\tau/\tau_R$ for the simulations depicted in Figs.~\ref{fig: figbr1} and \ref{fig: figbr2}. For $\eta/s = 1$ and $B \sim 100$ fm$^{-2}$, $\varphi$ only exceeds 0.5 at the early stages of the evolution. Consequently, we observe only subtle hints of oscillations in this scenario. In contrast, when $\eta/s = 10$ and $B \sim 100$ fm$^{-2}$, $\varphi$ remains significantly above 0.5 during almost all the time evolution. As a result, this leads to the prominent oscillations observed for the transverse component of the shear-stress tensor in Fig.~\ref{fig: figbr2}.

\begin{figure}[H]
    \centering
    \subfigure[\ Longitudinal component]{\includegraphics[width=0.4\linewidth]{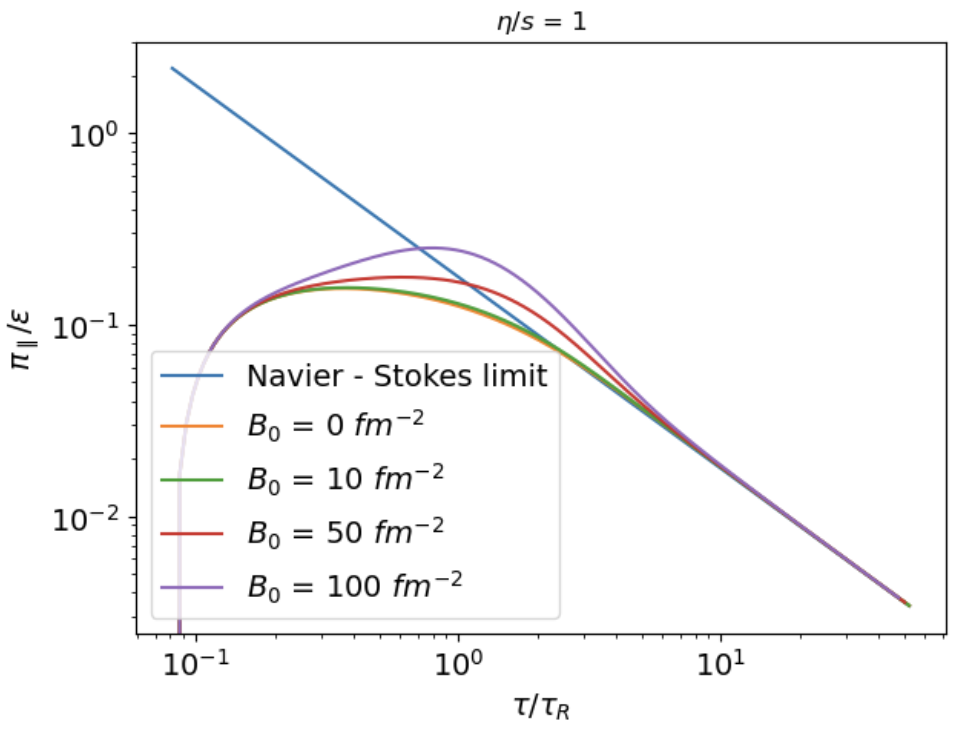}}
    \quad
    \subfigure[\ Transverse component ]{\includegraphics[width=0.4\linewidth]{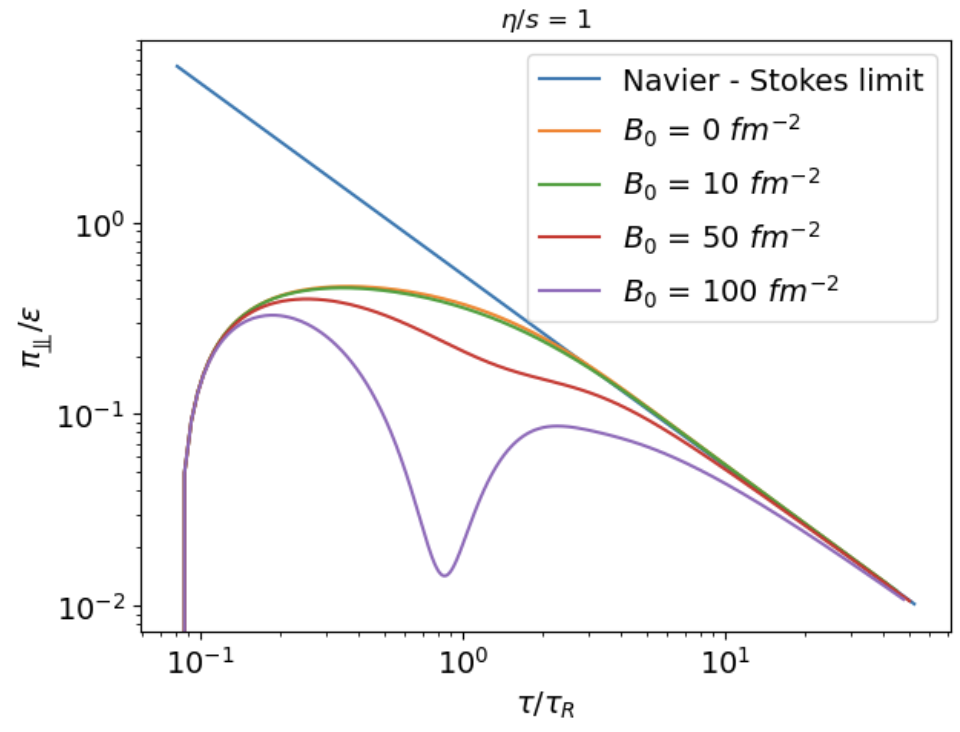}}
    \caption{Longitudinal and transverse components as function of $\tau/\tau_R$, compared to their respective Navier-Stokes values, for $\eta/s = 1$ and several values of $B_0$.}
    \label{fig: fignav}
\end{figure}

Due to the rapid decay of magnetic fields, the system is expected to ultimately approach the conventional Navier-Stokes limit at later times. In Fig.~\ref{fig: fignav} we confirm this behaviour by comparing our numerical solutions for the longitudinal (left panel) and transverse (right panel) components of the shear-stress tensor to their corresponding asymptotic Navier-Stokes values (blue line). For the sake of simplicity, we show our results only for the case of $\eta/s=1$. We note that that when the magnetic field is sufficiently large, the transverse components relax to their Navier-Stokes limit displaying oscillatory behaviour.

After examining this basic model in an expanding scenario, it becomes apparent that Israel-Stewart-like theories effectively work only when the quantity $\varphi$ is sufficiently small, with such threshold depending on the value of the shear viscosity of the gas. In other words, it is applicable primarily to lower magnetic field values that, in practical terms, exert little to no impact on the system's dynamics. For those wishing to investigate the effects of moderately stronger magnetic fields, it becomes necessary to resort to the fundamental coupled equations where no second-order truncation scheme has been implemented.

\section{Conclusions}
\label{sec:conclusion}
We derived the equations of motion of relativistic magnetohydrodynamics, as well as microscopic expressions for all of its transport coefficients, from the Boltzmann equation using the method of moments. In contrast to Refs.~\cite{denicol2018nonresistive, denicol2019resistive}, where a single component gas was considered, we perform our derivation for a locally neutral fluid composed of two particle species with opposite charges. For the sake of simplicity, the particles are assumed to be massless and have vanishing dipole moment or spin. Furthermore, we focus solely on understanding how the equations of motion for the shear-stress tensor are modified by the presence of a magnetic field.

The magnetohydrodynamic equations derived here are qualitatively different to the traditional Israel-Stewart theory. In our case, the longitudinal, semi-transverse and transverse components of the shear-stress tensor, with respect to the direction of the magnetic field, obey distinct equations of motion, with different relaxation times and viscosities. We have derived the microscopic expressions for these transport coefficients and shown how each of them is affected by the magnetic field. In particular, we found that the relaxation time appearing in the equations of motion for the longitudinal component of the shear-stress tensor displays no dependence on the magnetic field. On the other hand, the remaining relaxation times display a strong dependence on the magnetic field, being reduced as the magnetic field increases.    

We further demonstrate that the relaxation-type equations of motion derived for the semi-transverse and transverse components of $\pi^{\mu\nu}$ break down at sufficiently large magnetic fields, due to the fact that the relaxation times become negative. This departure from conventional expectations, with occurrence of negative relaxation times, challenges the applicability of standard relaxation-type equations to our system. For instance, we demonstrate that, when the magnetic oscillation frequency is of the same order as the inverse relaxation time, the system exhibits intrinsic oscillatory dynamics as it approaches its asymptotic Navier-Stokes regime and such behavior can never be described with relaxation-type equations of motion. A similar behavior is also observed for the transient dynamics of conformal fluids described using holography via the AdS/CFT correspondence \cite{Heller:2014wfa}. 

For the simple two-component gas considered in this work, it was possible to describe this novel oscillatory dynamics by including one additional dynamical variable: the dissipative dynamics of the fluid was described in terms of the total shear-stress tensor ($\pi^{\mu\nu}=\pi^{\mu\nu}_{+}+\pi^{\mu\nu}_{-}$) and its relative value ($\delta\pi^{\mu\nu}=\pi^{\mu\nu}_{+}-\pi^{\mu\nu}_{-}$). Nevertheless, it is not clear at this point how this oscillatory dynamics can be described when more complicated systems, with more particle species and realistic cross sections, are considered. These issues will be discussed in future works, where we shall also include the effects of electric conductivity, finite electric chemical potential and finite particle masses.

Our work addressed some of the theoretical challenges in deriving relativistic magnetohydrodynamics and motivates for further exploration. Our analyses into the nuanced behavior of magnetohydrodynamics in stronger magnetic fields, suggests the need for more comprehensive theoretical frameworks to capture the dynamics of relativistic fluids in the presence of powerful magnetic fields. In particular, we advocate for the development of a more robust theoretical framework capable of capturing and incorporating the oscillatory dynamics that appear at strong magnetic fields, even for more general fluids.

\section*{Acknowledgements}
K.~Kushwah and G.~S.~Denicol thank H.~Niemi, A.~Jaiswal and Sourav Dey for insightful discussions. The authors acknowledge Conselho Nacional de Desenvolvimento Científico e Tecnológico (CNPq) and A Fundação Carlos Chagas Filho de Amparo à Pesquisa do Estado do Rio de Janeiro (FAPERJ) for financial support.

\bibliographystyle{apsrev4-1}
\bibliography{references.bib}
\end{document}